\documentclass[iop]{emulateapj}
\usepackage{mathptmx}
\usepackage{natbib}
\usepackage{units}
\usepackage{graphicx}
\usepackage{url}
 \usepackage{color}
\bibpunct{(}{)}{;}{a}{}{,}
\usepackage[colorlinks=true, urlcolor=blue, citecolor=blue, linkcolor=blue]{hyperref}

\newcommand{\dd}{\ensuremath{\mathrm{d}}}

\newcommand{\be}{\begin{equation}}
\newcommand{\ee}{\end{equation}}

\newcommand{\bifrost}{{\textsl{Bifrost}}}

\newcommand{\ci}{\ion{C}{1}}
\newcommand{\cii}{\ion{C}{2}}

\newcommand{\mgii}{\ion{Mg}{2}}
\newcommand{\mgiik}{\mgii~k}
\newcommand{\siiv}{\ion{Si}{4}}

\newcommand{\ie}{{\it i.e.,}} 
\newcommand{\eg}{{\it e.g.,}} 
\newcommand{\is}{\ensuremath{\!=\!}}
\newcommand{\ciib}{\ion{C}{2}~\ensuremath{133.5\,{\rm nm}}}
\newcommand{\ciia}{\ion{C}{2}~\ensuremath{133.4\,{\rm nm}}}
\def\edt#1{{#1}}

\begin{document}
\title{%
 The formation of  {\it IRIS} diagnostics
 \\ VI. The Diagnostic Potential of the  \mbox{C II} Lines at $133.5\,$nm in the Solar Atmosphere
}%
\author{Bhavna Rathore\altaffilmark{1}}
\author{Mats Carlsson\altaffilmark{1}}
\author{ Jorrit Leenaarts\altaffilmark{2}}
\author{ \and Bart De Pontieu\altaffilmark{3,1}}

\affil{
{\altaffilmark{1}Institute of Theoretical Astrophysics, University of Oslo, P.O. Box 1029 Blindern, N-0315 Oslo, Norway}\\
{\altaffilmark{2}The Institute for Solar Physics, AlbaNova University Center, SE-106 91 Stockholm, Sweden}\\
{\altaffilmark{3}Lockheed Martin Solar \& Astrophysics Lab, Org. A021S, Bldg. 252, 3251 Hanover Street Palo Alto, CA 94304 USA}
}
\email{bhavna.rathore@astro.uio.no}
\email{mats.carlsson@astro.uio.no}
\email{jorrit.leenaarts@astro.su.se}
\email{bdp@lmsal.com}
\begin{abstract}
We use 3D radiation magnetohydrodynamic models to investigate how the thermodynamic quantities in the simulation are
encoded in observable quantities, thus exploring the diagnostic potential of the \ciib\ lines.
We find that the line core intensity is correlated with the temperature at the formation height
but the correlation is rather weak, especially when the lines are strong.
The line core Doppler shift is a good measure of the line-of-sight velocity at the formation height. 
The line width is both dependent on the width of the absorption profile (thermal and non-thermal width) and an
opacity broadening factor of 1.2-4 due to the optically thick line formation with a larger broadening for double
peak profiles.
The \ciib\ lines can be formed both higher and lower than the core of the \mgiik\ line depending on the amount
of plasma in the 14--50~kK temperature range. 
More plasma in this temperature range gives a higher \ciib\ formation height relative to the \mgiik\ line core.
The synthetic line profiles have been compared with IRIS observations. The derived parameters from the simulated
line profiles cover the parameter range seen in observations but on average the synthetic profiles are too narrow. 
We interpret this discrepancy as a combination of a lack of plasma at chromospheric temperatures in the simulation box and too small non-thermal velocities. 
The large differences in the distribution of properties between the synthetic profiles
and the observed ones show that the \ciib\ lines are powerful diagnostics of the upper chromosphere and 
lower transition region.

\end{abstract}


\section{Introduction}

This is the second paper in the series exploring the 
diagnostic potential of the \cii\ lines around $133.5\,$nm under solar conditions. 
There are three components in the multiplet: at $133.4532\,$nm, hereafter called the \ciia\ line,
and at $133.5708\,$nm with a weaker blend at $133.5663\,$nm, hereafter together called the \ciib\ line. 
These lines are among the strongest lines in the far ultraviolet (FUV) range of the NASA's {\it Interface
Region Imaging Spectrograph} ({\it IRIS}) space mission and exploring their diagnostic potential is therefore of 
great interest.
In \citet{cii_paper1} (hereafter called Paper~I) we analysed the basic rate balance under solar chromospheric
conditions and showed that a nine-level model atom sufficed to describe the ionization and excitation balance
for the proper modelling of the \cii\ lines around $133.5\,$nm.
We also studied the general formation processes in Paper~I and found that the lines are mainly formed in the
optically thick regime with an average formation temperature of 10~kK and a line core formation close to 
a column mass of $10^{-6}$ g~cm$^{-2}$.

For references to earlier work on these \cii\ lines we refer to Paper~I. We will here focus on the diagnostic
potential of the \cii\ lines and explore correlations between observable quantities and properties of a 
snapshot from a 3D MHD model calculated with the \bifrost\ code \citep{2011A&A...531A.154G}. We will
also compare the synthetic observables with observations from {\it IRIS} to determine the applicability
of the deduced relations.

The organization of this paper is as follows: 
In Section~\ref{sec:rt} we describe the radiative transfer code used, 
in Section~\ref{sec:ma} we present the snapshot from the 3D MHD simulation,
in Section~\ref{sec:basic_characteristics} we present the synthetic spectra and discuss line profiles, how to define the
line core and formation heights. 
In Section~\ref{sec:diagnostics} we discuss the diagnostic potential of the lines.
We compare with other spectral lines in the {\it IRIS} passbands in Section~\ref{sec:compare_iris}
and with {\it IRIS} observations in Section~\ref{sec:observations}.
We summarize and add concluding remarks in Section~\ref{sec:discussion}.

\section{Radiative Transfer}\label{sec:rt}
In the present work, we have used 
the full 3D radiative transfer code MULTI3D \citep{2009ASPC..415...87L}. We have used the same 9-level 
quintessential atomic model for \cii\ arrived at in Paper~I. The MULTI3D code solves the coupled 
statistical equilibrium and radiative transfer equations in full 3D. For our computations we assume complete
frequency redistribution (CRD), an approximation that was shown to be adequate in Paper~I. 
The blend between the 
$133.5708\,$nm and $133.5663\,$nm lines is taken into account self-consistently as are the overlaps
between the \ci\ photoionisation continua \citep{1991A&A...245..171R,1992A&A...262..209R}.
The code includes the local approximate operator of \edt{\citet{1986JQSRT..35..431O}}, Ng acceleration 
\citep{1974JChPh..61.2680N}, collisional-radiative switching  \citep{1988A&A...192..279H} and  background opacities 
from the Uppsala Opacity Package \citep{Gustafsson1973}.

 \section{Model atmosphere}\label{sec:ma}
 
To explore relations between observable quantities and atmospheric quantities we have used a 3D
snapshot from the simulation {\it en024048\_hion} calculated with the 3D radiative magnetohydrodynamic (RMHD) code
\bifrost, \citep{2011A&A...531A.154G}. We use the simulation snapshot 385, the same as has been used in the
previous papers on the formation of {\it IRIS} diagnostics
\citep{2013ApJ...772...89L, 2013ApJ...772...90L,2013ApJ...778..143P,2015arXiv150401733P} and a number of other papers on 
line formation under solar chromospheric conditions:
\citep{2012ApJ...749..136L, 2012ApJ...758L..43S, 2013ApJ...764L..11D}. 
The full simulation cubes with all variables as function of grid
position are available from the European Hinode Science Data Centre
(http://www.sdc.uio.no/search/simulations). A detailed description of the simulation {\it en024048\_hion}
is given in \citet{chromsim15} and we summarise some of the properties below.

The simulation box is $24\! \times \! 24 \! \times \!16.8\! $ Mm$^3$ discretised onto 
$504\! \times \! 504\! \times \! 496$ grid points. The vertical extent is from $2.4$ Mm below to $14.4$ Mm above
the photosphere and covers the upper convection zone, photosphere, chromosphere and lower corona. 
The horizontal grid is equidistant with 48 km spacing and periodic boundaries. Vertically, the grid spacing 
is 19 km from $z\is -1$~Mm up to $z\is 5$~Mm. The spacing increases towards the lower and upper boundaries
to a maximum of 98 km at the coronal boundary. The magnetic field was introduced into the computational box 
as two opposite polarity blobs separated by 8~Mm. This large scale structure is evident throughout the 
simulation although a lot of small scale structure develops from the action of the convection on the magnetic field. 
The mean unsigned field strength is 50~G in the photosphere. 

\begin{figure}[hbtp]
  \centering
  \includegraphics[width=\columnwidth]{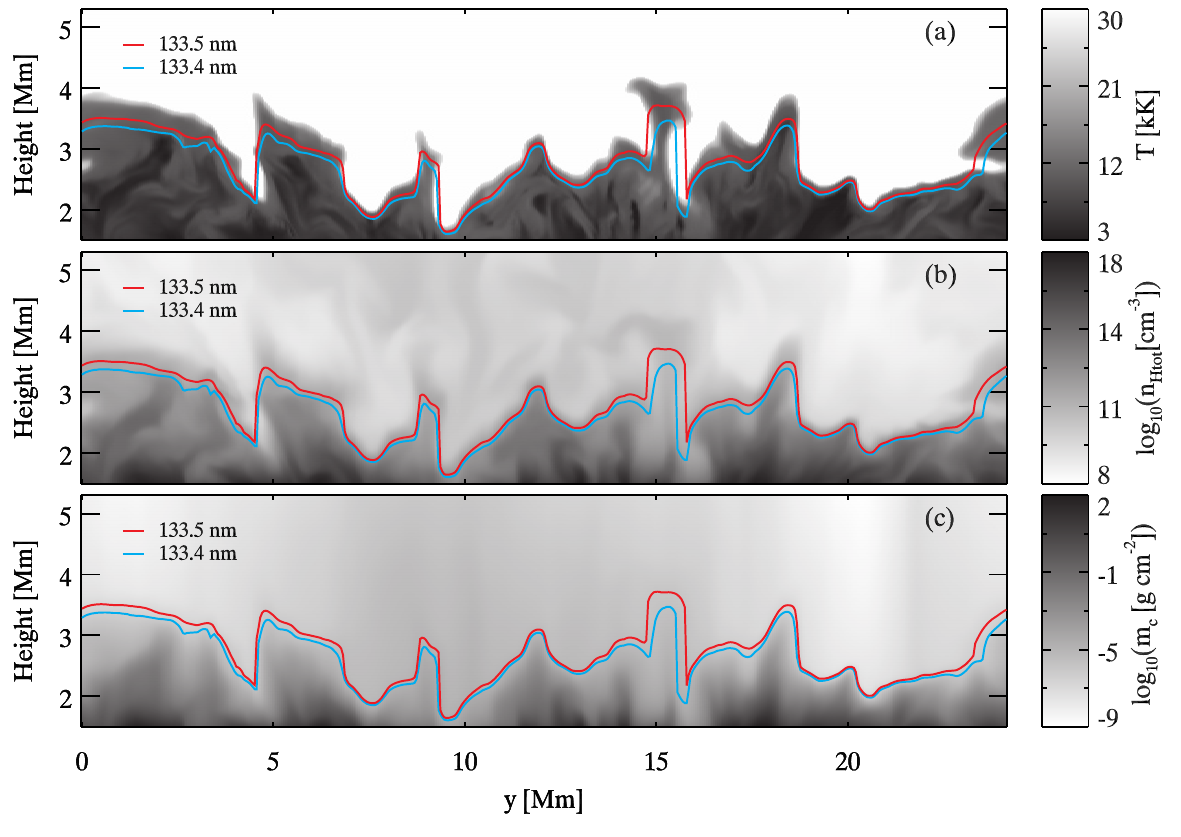}
  \caption{\label{t_nh_cm}
Temperature (a) , total hydrogen population density (b), and column mass (c) as function of height 
along a 2D cutout at $x\is 12\,$Mm from the 3D model atmosphere.
The maximum height across the line profile of $\tau_\nu=1$ is shown for the  \ciib\ (red) and 
\ciia\ (blue) lines.}
\end{figure}

We illustrate the simulation box by showing a cut at $x\is 12\,$Mm in Figure~\ref{t_nh_cm}. The figure shows
the temperature, 
the total hydrogen population density and 
the column mass as function of height and position along
the y-coordinate in the simulation box. 
The figure shows \edt{the} line core
\edt{height of unit optical depth} of both the \cii\ lines as red and blue lines. 
At this line core formation height, we have a temperature in the range of $9\!-\!14 $ kK, 
the density is of the order of $10^{-14} \!$  to $10^{-12}\!$ g cm$^{-3}$, 
the total hydrogen particle density is $10^{9} \!$  to $10^{12}\!$  cm$^{-3}$ and
the column mass is of order  $10^{-6} \!$ g cm$^{-2}$. 

\section{Basic characteristics}\label{sec:basic_characteristics}
\subsection{Line profiles}
\begin{figure}[hbtp]
  \includegraphics[width=\columnwidth]{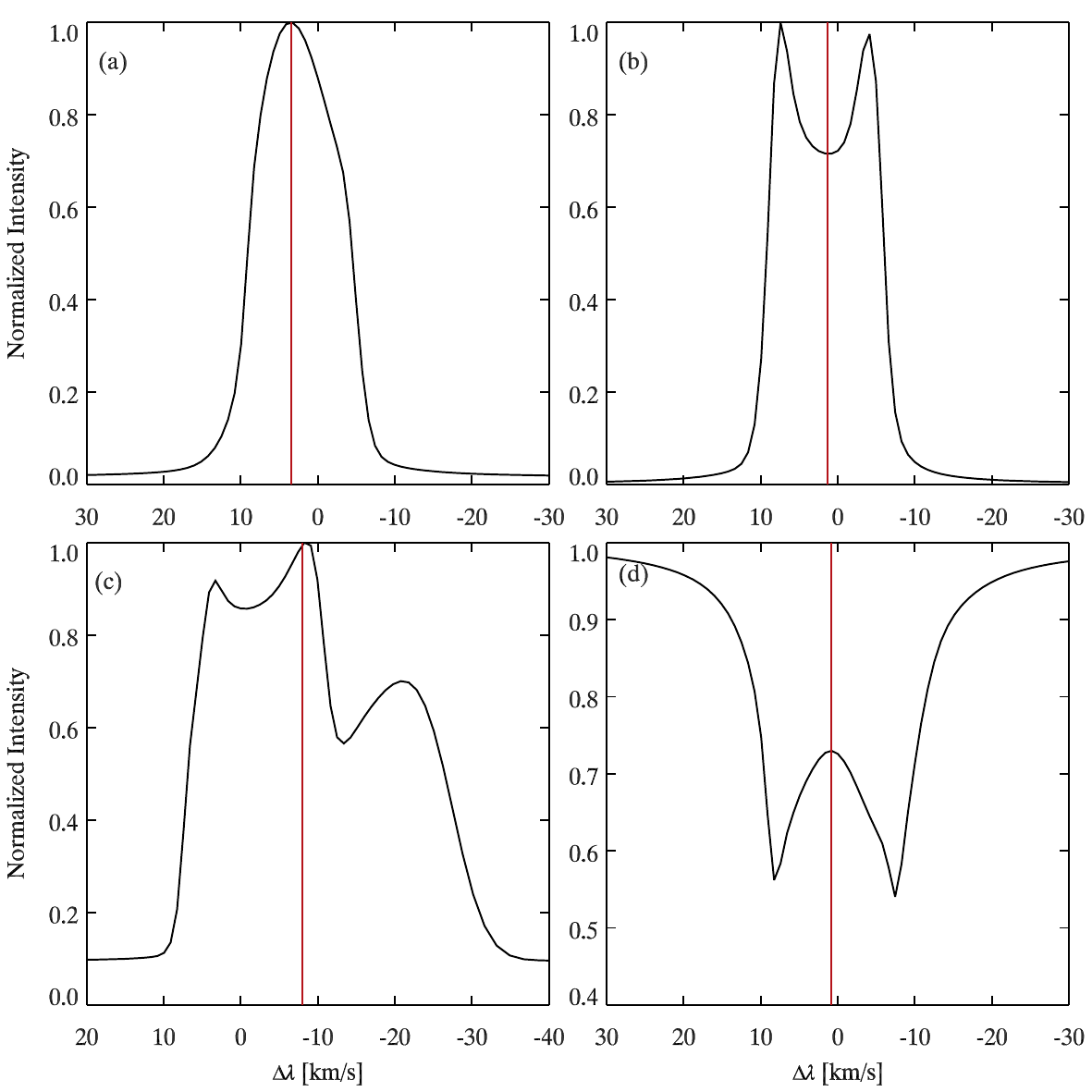}
  \caption[]{\label{lc} 
Various \cii \ line profiles from our simulation with the position of the line center using our peak finding algorithm
shown with a red vertical line: typical single-peak emission line profile (a), double-peak profile with central depression (b), profile
showing three peaks (c), absorption line profile (d). }
\end{figure}

From this simulation we get a very broad range of intensities and shapes of the line profiles. The line profiles
can have a single peak, double peaks, three peaks and even be in absorption. Some typical line profiles
are  shown in Figure~\ref{lc}. 
We show these line profiles and make all our subsequent analysis at the full resolution of the simulation, without smearing
the profiles to {\it IRIS} resolution. 
We will discuss the effects of the {\it IRIS} resolution on the results in Section~\ref{sec:discussion}.

\edt{The diversity of intensity profiles is the result of the interplay between the source function and the
optical depth variation along the line of sight. \edt{S}ingle-peak profiles result from source functions that increase
monotonically with height up to the height where the line core has optical depth unity. We get a double peak
when the source function has a local maximum and then decreases before the height of optical depth unity 
in the line core. We may get 
more peaks when we have source functions with multiple local maxima. This scenario is further complicated
by velocity gradients that Doppler shift the maximum of the local absorption profile. These velocity gradients
may partly smear out an otherwise symmetric double-peak profile rendering it single-peak, but asymmetric, with the intensity
maximum on the red or blue side of the line core. }

\begin{figure}[hbtp]
  \centering
  \includegraphics[width=\columnwidth]{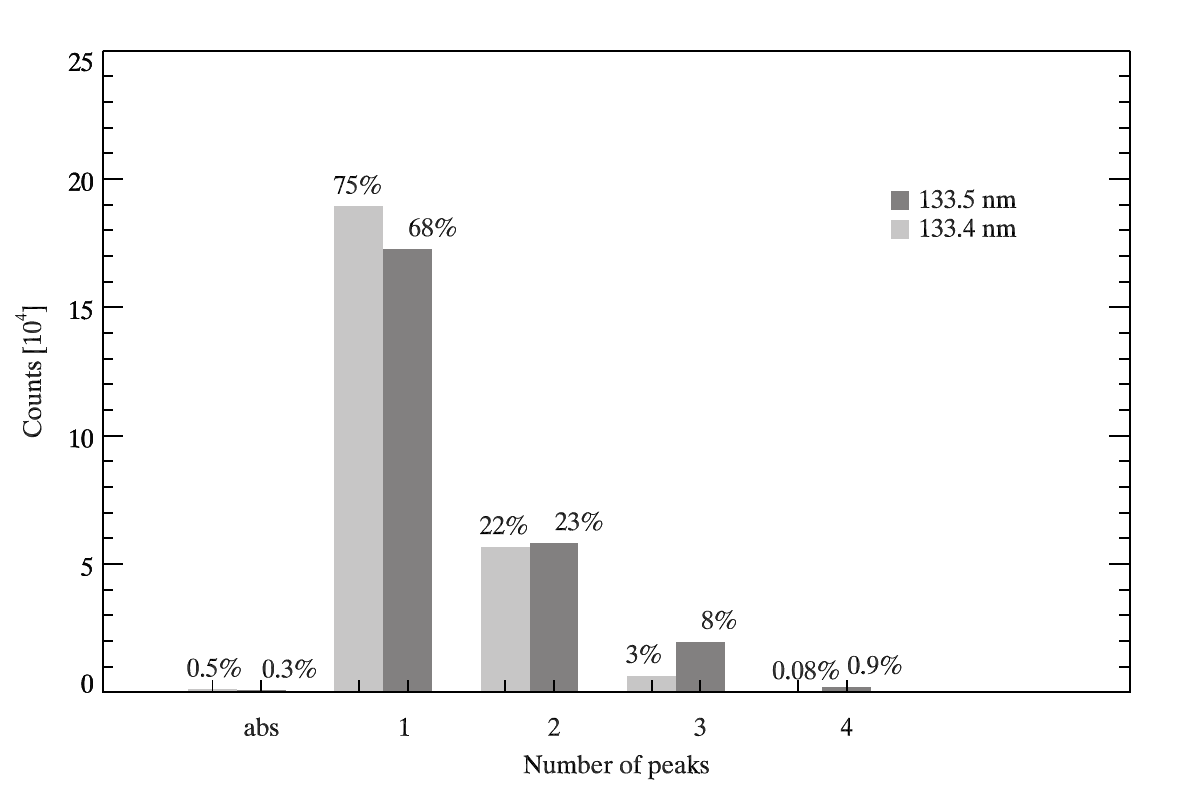} 
  \caption[]{\label{npeak} 
Statistics of number of intensity peaks in the  \ciia\ and \ciib\ lines in the 3D snapshot. }
\end{figure}

\edt{The number of peaks varies between the lines. The \ciia\ line has more single-peak profiles than the
\ciib\ line (see Figure~\ref{npeak}) . The explanation is that the \ciib\ is the stronger of the two
main components of the multiplet so the optical depth unity point is higher up in the atmosphere
(see also Figure~\ref{t_nh_cm}). There is thus a higher probability that the line core is formed above
the height where the source function has a local maximum, thus causing a central reversal (\ie\ double-peak profile)
more easily for the \ciib\ line. }
 
\subsection{Defining the line core}\label{sec:linecore}
Diagnostics derived from a given wavelength in the line, \eg\ the line core, can be expected to give information
from a narrow region of the atmosphere compared with diagnostics derived from the full line profile, \eg\ moments
of the intensity or parameters from a Gaussian fit. 
In Paper~I we defined the line core as the wavelength where the height at which optical depth reaches unity is maximal.
This is a reasonable definition from a theoretical point of view but it is not practical observationally since the
optical depths are not known. 

In an attempt to get a good observational definition of the line core we have made an algorithm to pick the 
position of the line core for all kinds of line profiles. To extract the line center position we consider
the profile within the spectral range of ($-50 < \Delta \rm{v} < 50 $ km s$^{-1} $) around the rest wavelength
of each line. We define the line core on the basis of the number of maxima and minima in the line profile. 
For single-peak profiles, the line core is simply defined to be at the central maximum. 
For double-peak profiles the line core is defined to be at the minimum between the two maxima. 
For three peak profiles the line core is defined to be at the maximum between the two minima. 
Similarly, for an even number of peaks the line core is defined to be at the central minimum and for an odd number of
peaks, the line core is defined to be at the central maximum. 
A similar procedure is adopted for absorption profiles. Line centers according to this algorithm 
are shown in Figure~\ref{lc}. 
To avoid small peaks close to the continuum intensity, we only consider the line profile with intensity 50\% above
the continuum. For real observations, care must be taken not to count maxima caused by noisy data. This can be
achieved by first applying smoothing or noise filtering (\eg\ Wiener filtering) to the data.

\begin{figure}[hbtp]
  \centering
  \includegraphics[width=\columnwidth]{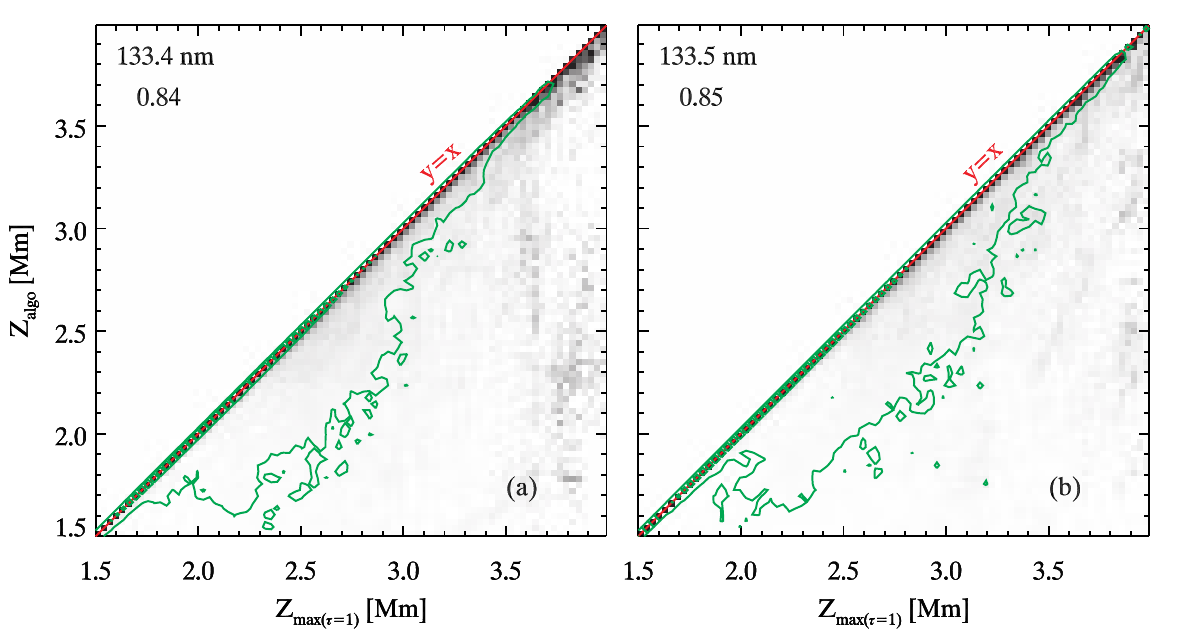}
  \caption[]{\label{zfm_lc_algo_pdf}
Probability distribution function (PDF) of $\tau =1$ height at the line center determined by the line center finding algorithm as a function
of the theoretical line core $\tau =1$ height (the maximum $\tau = 1$ height over the line profile) 
for the \ciia\ line (a) and the \ciib\ line (b). 
Each column in the panels is scaled to maximum contrast to increase visibility.
The inner green contour encompasses $50\%$ and the outer contour encompasses $90\%$ of all points. 
The red line denotes y=x. 
The Pearson correlation coefficient is given in the upper left corner.
}
\end{figure}

Figure~\ref{zfm_lc_algo_pdf} shows the relation between the theoretical line core optical depth unity
height (the maximum $\tau_\nu\is 1$ height over the line profile) and the height of $\tau_\nu\is 1$ at
the wavelength of the line center determined with our algorithm. For a majority of the profiles, the 
algorithm works well in finding the same wavelength as the theoretical line core but there are some
cases where the observationally found line center is quite different from the theoretical one, resulting
in a large difference between the heights. 
This may happen when effects of large velocity gradients smear 
out local minima such that a profile that would exhibit two peaks in the absence of velocity fields appears
single peaked and rather asymmetric. Our algorithm then picks out the intensity maximum, which
can be some distance away from the theoretical line core. For 75\% of the columns in the 3D model, the
difference between the two heights is less than 100~km while for 5\% of the columns, the difference is
larger than 1~Mm.

\subsection{Formation height}\label{sec:formation_height}
For optically thick line formation, the Eddington-Barbier relation often gives a good indication of 
where the intensity comes from. The Eddington-Barbier relation states that the disk center outgoing intensity is
approximately equal to the source function at optical depth unity. This relation is exact if the source
function is a linear function of optical depth but is often a good approximation also when the source
function is quite non-linear (which is normally the case in the ultraviolet part of the spectrum where
the Planck function is close to an exponential function of temperature).

\begin{figure}[hbtp]
  \centering
  \includegraphics[width=\columnwidth]{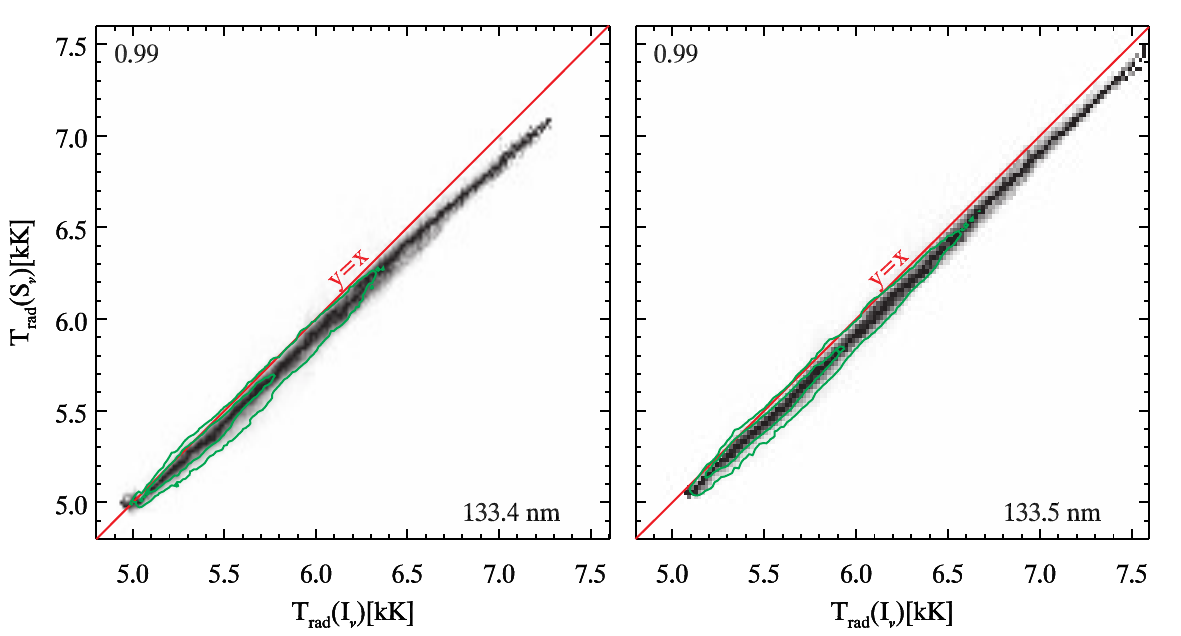}
  \caption[]{\label{edd} 
PDF of the radiation temperature of the source function at the $\tau_\nu\is 1$ 
height as function of the radiation temperature of the intensity. The correlation is shown for the theoretical line
core (the wavelength where the height where optical depth unity is the maximum).
Each column in the panels is scaled to maximum contrast to increase visibility.
The green contours encompass 50\% and 90\% of all points. 
The Pearson correlation coefficient is given in the upper left corner, the red line denotes $y\is x$.}
\end{figure}

Figure~\ref{edd} shows the correlation between the line intensity at the theoretical line core (\edt{see Section~\ref{sec:linecore}})
and the source function at  the $\tau_\nu=1$ height.
The correlation is quite good, especially for the lower intensity values. The higher intensities correspond to
columns where the source function at optical depth unity is high. These columns often show a steep source
function rise with height and therefore an average contribution shifted towards \edt{optical depths smaller than unity} where
the source function is higher.

Figure~\ref{edd} tells us that optical depth unity is 
a good first approximation to the formation height except 
for columns with a steep, non-linear, temperature gradient where optical depth unity gives too low heights (and too low formation temperatures).
Instead of using a one-point integration formula for the relation between the source function and the
outgoing intensity (like the Eddington-Barbier relation), we may use the formal solution to the transfer 
equation to define the
contribution function to intensity at solar disk center, $C_{I_\nu}(z)$:
\begin{equation}\label{ci}
C_{I_{ \nu}}(z)=S_{\nu}(z) e^{-\tau_{\nu}(z)} \chi_{\nu}(z),
\end{equation}
where $S_{\nu}$ is the source function, $\chi_{\nu}$ is the extinction coefficient and $\tau_{\nu}$ is the optical 
depth at frequency $\nu$, with all these quantities functions of height, $z$. 

The \cii\ lines are mainly formed in the optically thick regime. However, we may have a substantial optically 
thin component to the intensity in the cases where the source function rises rapidly with height, as 
seen in Figure~\ref{edd}. There 
are also cases among the low-intensity profiles where this is the case, as also exemplified in Paper~I.
To further quantify the relative importance of the optically thin and thick contributions, we use the
contribution function to the total intensity: 
\begin{equation}\label{ci_tot}
C_{I_{\rm total}}(z) =\int C_{I_\nu}(z) \, \dd\nu
\end{equation}
The contribution to total intensity on a $\lg \tau_0$ scale ($\lg$ denotes the logarithm to the base of 10) is
obtained from
\begin{equation}\label{ci_log}
C_{I_{\rm total}}(\lg \tau_0) =   C_{I_{\rm total}}(z) \frac{\dd z}{\dd \lg \tau_0}
\end{equation}
where $\tau_0$ is the optical depth at the line core.

We also define the mean formation depth in $\lg \tau_0$ from
\begin{equation}\label{tfm}
\lg \tau_{\rm fm} = \frac {\int   \lg \tau_0 \, C_{I_{\rm total}}(\lg\tau_0)  \, \dd \lg\tau_0}{\int C_{I_{\rm total}}(\lg{\tau_0})  
\dd \lg\tau_0}
\end{equation}

\begin{figure}[hbtp]
  \centering
  \includegraphics[width=\columnwidth]{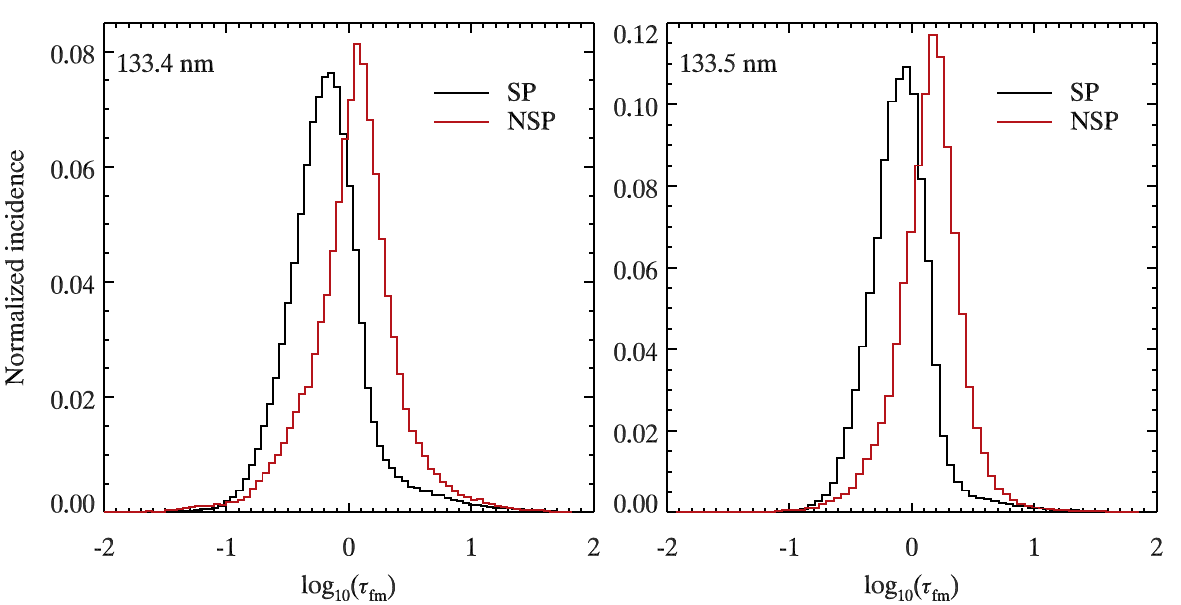} 
  \caption[]{\label{lgtau_hist}
Histogram of average formation height on a $\lg(\tau_0)$ scale (see Equation~\ref{tfm}) for single-peak (SP) 
(black) and 
non-single-peak (NSP) (red) profiles for the \ciia\ line (left) and the \ciib\  line (right).}
\end{figure}

If we have a substantial thin contribution to the total intensity, we expect the mean formation depth to be
at an optical depth significantly smaller than one. Figure~\ref{lgtau_hist} shows a histogram of the
formation depth in $\lg \tau_0$ for all the columns in the simulation box. We show histograms separately 
for single-peak profiles and the other profiles. For non-single-peak profiles we have a distribution that is
centered on $\lg \tau_{\rm fm}=0-0.2$ while single-peak profiles show distributions skewed to smaller optical
depths. This clearly shows the effect of a steep non-linear source function rise towards smaller optical depths. 
However, there are very few columns where the formation could be described as being dominated by
a very thin contribution.

\begin{figure}[hbtp]
  \centering
  \includegraphics[width=\columnwidth]{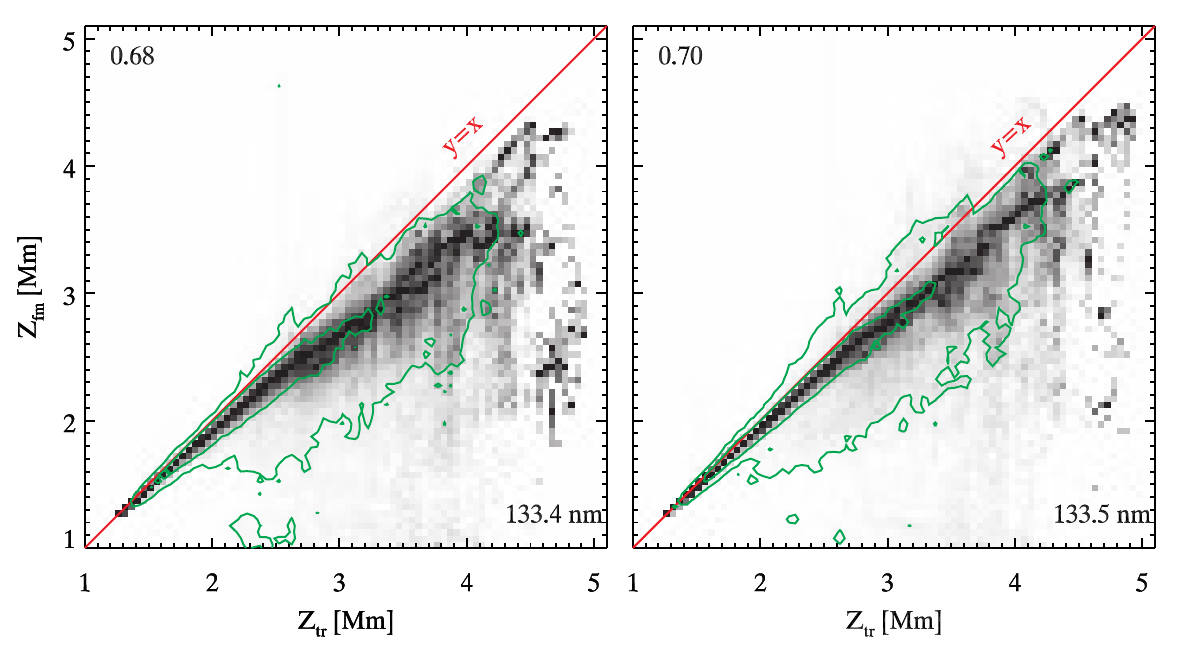}
  \caption[]{\label{corr_ztr}
PDF of formation height as a function of transition region height (defined to be the lowest height in 
a column where the temperature is greater than 30~kK) for the \ciia\ line (left) and the 
\ciib\ line (right). 
Each column in the panels is scaled to maximum contrast to increase visibility.
The green contours encompass 50\% and 90\% of all points. 
The Pearson correlation coefficient is given in the upper left corner, the red line denotes y=x.}
\end{figure}

We now turn to describing the formation height relative to the height of the transition region.
We define the formation height $Z_{\rm fm}$ from the contribution function weighted average height:
\begin{equation}\label{zfm}
Z_{\rm fm}(\nu)=\frac{\int z \ C_{I_{\nu}}(z)  \ {\rm d} z} {\int C_{I_{\nu}}(z) \ {\rm d} z}.
\end{equation}
Figure~\ref{corr_ztr} shows the relation between the transition region height and this formation height
for the theoretical line core wavelength for the \ciia\ and \ciib\ lines. The transition region height is
defined as the lowest height where the temperature is greater than 30~kK. 
Both lines are formed just below this transition region height. The difference in height is larger when the
transition region is higher in the atmosphere. The formation height is very close to the transition region
when the transition region is very low in the atmosphere. This corresponds to locations of strong flux
concentrations in the photosphere.

\section{Diagnostic potential}\label{sec:diagnostics}

In Paper~I and in Section~\ref{sec:basic_characteristics} we established the basic formation characteristics of the
\cii\ lines. In this section we will correlate observables with the hydrodynamical state in the simulation in order
to explore the diagnostic potential of the \cii\ lines. The figures are organised with the observable quantity
along the x-axis and the atmospheric property along the y-axis. We normally show the Probability Density Function
(PDF)  of the atmospheric quantity as a function of the observed quantity 
normalized by column to bring out the correlation also where there are few points. 
What appears as a bad correlation at the extreme ranges may thus affect only a small number of cases. 
Contours are therefore added that encompass 50\% and 90\% of the points in the simulation.
We start by looking at the intensity, continue with line
shifts and finish by looking at the line widths.

\subsection{Intensity}\label{sec:intensity}

\subsubsection{vs Formation height}

\begin{figure}[hbtp]
  \centering
  \includegraphics[width=\columnwidth]{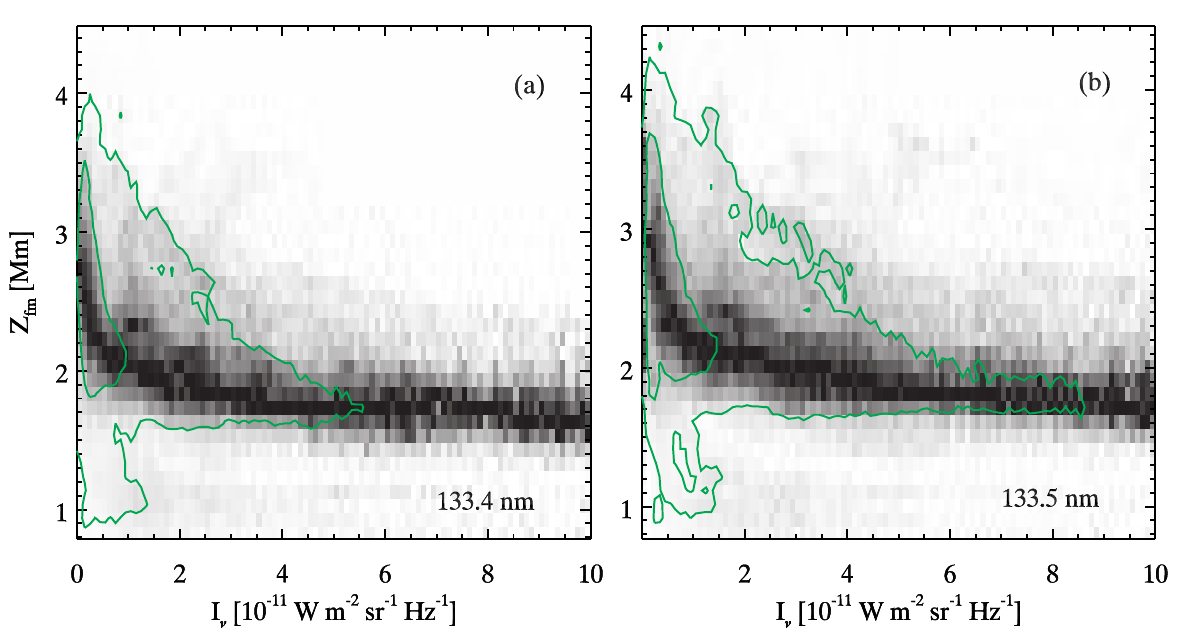} 
   \caption[]{\label{int_zfm}
PDF of formation height (see Section~\ref{sec:formation_height} and Equation~\ref{zfm}) as function of intensity 
at the line core for \ciia\ (left panel) and \ciib\ (right panel). 
Each column in the panels is scaled to maximum contrast to increase visibility.
The green contours encompass 50\% and 90\% of all points. 
}
\end{figure}

Figure~\ref{int_zfm} shows the  correlation between line core intensity and formation height. The
low intensity profiles tend to have a higher formation height. 
\edt{Since low intensity means low source function at the formation height (Figure~\ref{edd}) and the lines
are formed close to the transition region (Figure~\ref{corr_ztr}), 
the relation shown in Figure~\ref{int_zfm} thus means that the 
source function is low when the transition region is at a large height.}

\subsubsection{vs Temperature}
\begin{figure}[hbtp]
  \centering
  \includegraphics[width=\columnwidth]{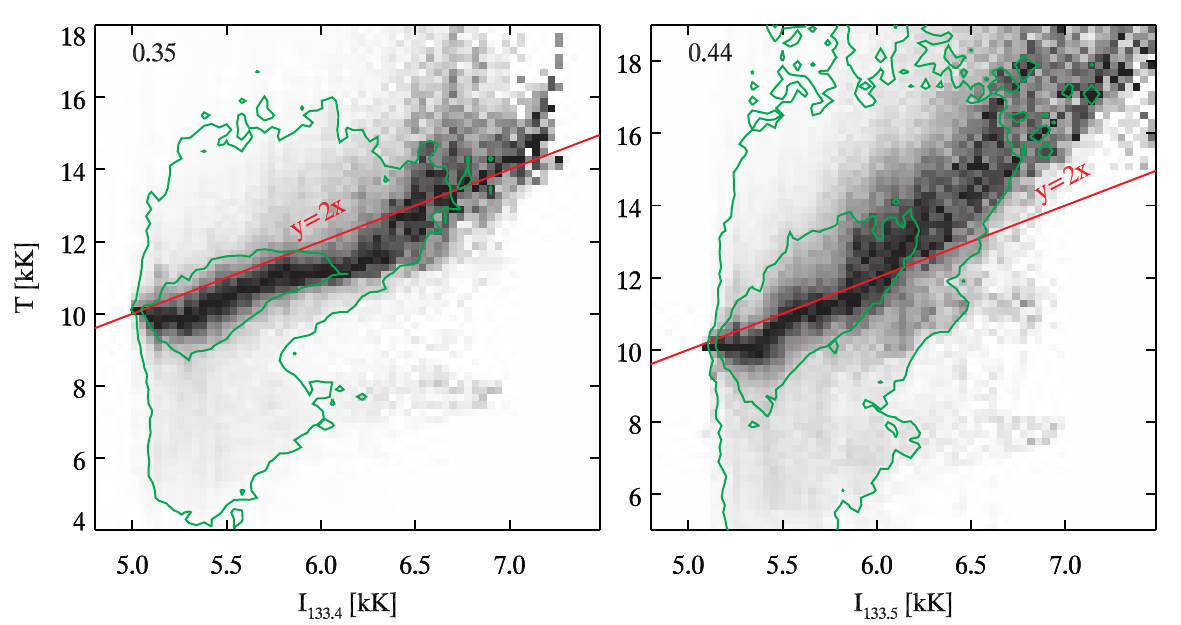}  
   \caption[]{\label{int_temp}
PDF of the temperature at the formation height as a function of the radiation temperature of the intensity at the 
line core for the \ciia\ line (left panel) and the \ciib\ line (right panel). 
Each column in the panels is scaled to maximum contrast to increase visibility.
The green contours encompass 50\% and 90\% of all points. 
The Pearson correlation coefficient is given in the upper left corner of both panels. 
The red line denotes the line y=2x. }
\end{figure}

The line core intensity is closely correlated with the source function at optical depth unity 
(Figure~\ref{edd}) but how does the source function relate to the temperature? With
a close coupling between the Planck function and the source function, we could use the intensity as a probe
of the temperature at the height of formation.
Figure~\ref{int_temp} shows the correlation between the radiation temperature of the line core intensity 
and the temperature at the formation height. 
We find that the temperature at the formation height of the line is approximately twice the radiation temperature 
of the line core intensity of both the lines. 
This means that the source function has decoupled from the
Planck function at the formation height and the amount of decoupling varies quite a bit, causing a large
scatter and a small correlation coefficient. Inspection of the corresponding images of radiation temperature of
the core intensity and temperature at the formation height also shows that the spread is too large for the weak
correlation to be of practical use.

\subsubsection{Line ratio}\label{sec:lineratio}

\begin{figure}[hbtp]
  \centering
  \includegraphics[width=\columnwidth]{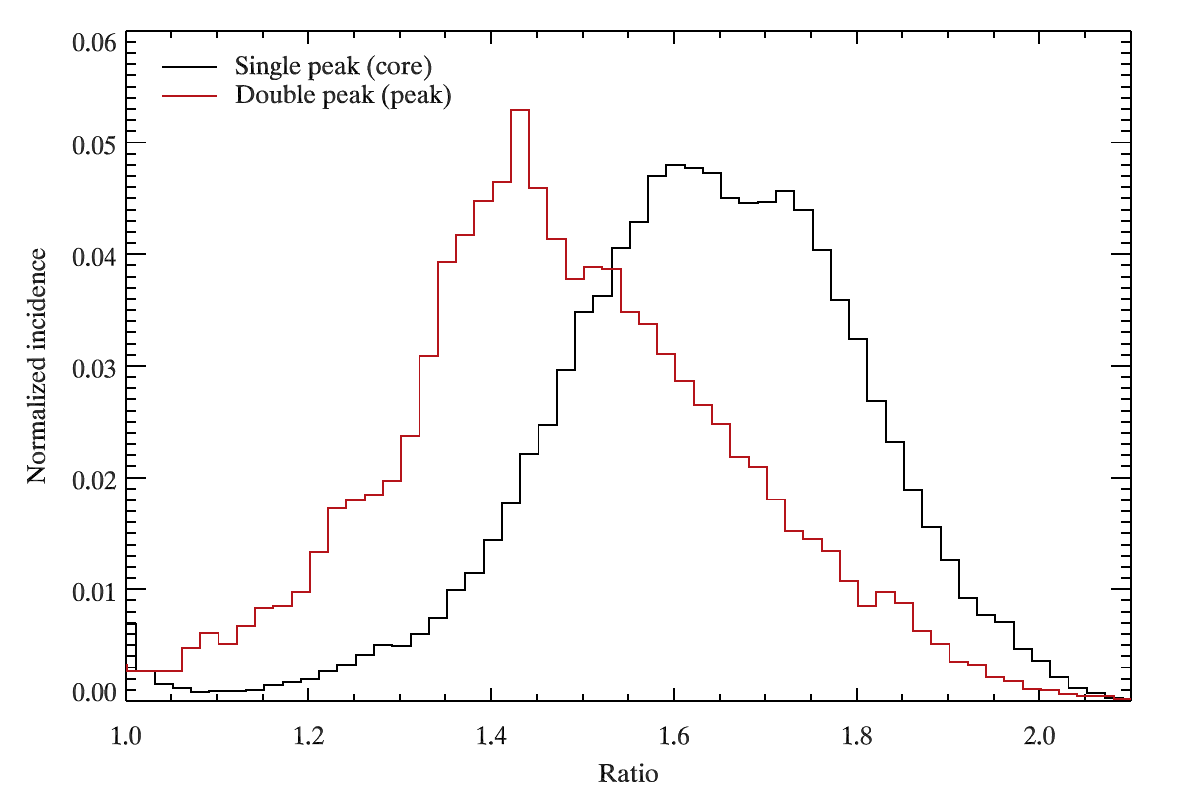} 
  \caption[]{\label{fig:ratio}
Histogram of the line peak intensity ratio of single-peak (black) and double-peak (red) profiles from the simulation.}
\end{figure}

The intensity ratio between the \ciib\ and the \ciia\ lines is 1.8 in the optically thin case and can be any value in the optically 
thick case depending on the ratio of the source functions of the two lines (see Paper~I). Figure~\ref{fig:ratio} shows a histogram
of this ratio in the \bifrost\ simulation.
The peak intensity ratio is around 1.7 for single-peak profiles and 1.4 for the peaks of double-peak profiles. 
The double peaks are formed lower in the atmosphere than the single peaks and because of the higher density, we expect the source functions to be more equal leading to a smaller ratio. 
In addition, the peak intensity for double-peak profiles is formed at the local source function maximum which is typically 
at the same height for both lines. We therefore have no effect of different formation heights for double-peak profiles,
while for single-peak profiles the \ciib\ intensity is formed higher in the atmosphere than the \ciia\ intensity. 
There the source function is higher, thus also leading to a higher line ratio.

\subsubsection{Line asymmetry}
 
Double-peak profiles often show an asymmetry with one peak brighter than the other. For the \mgii\ h \& k lines, 
\citet{2013ApJ...772...90L} showed that the asymmetry is caused by a velocity gradient between 
the heights of formation of the peaks and the line core. We define an asymmetry measure as
\begin{equation}\label{ir}
R_{c}=\frac{I_{b}-I_{r}}{I_{b}+I_{r}}
\end{equation}
where $I_{b}$ is the blue peak intensity and $I_{r}$ is the red peak intensity. 
$R_c$ is thus positive if the blue peak is stronger and negative if the red peak is stronger.

We furthermore define the velocity difference between the core and peak formation heights from

\begin{equation}\label{vd}
v_{\rm diff}=v(z_0) - v(z_p),
\end{equation}
where $v(z_p)$ is the velocity at the average of the heights of optical depth unity 
for the blue and the red peak and
$v(z_0)$ is the velocity at the height of optical depth unity for the line core. Positive velocity is upflow.

\begin{figure}[hbtp]
  \includegraphics[width=\columnwidth]{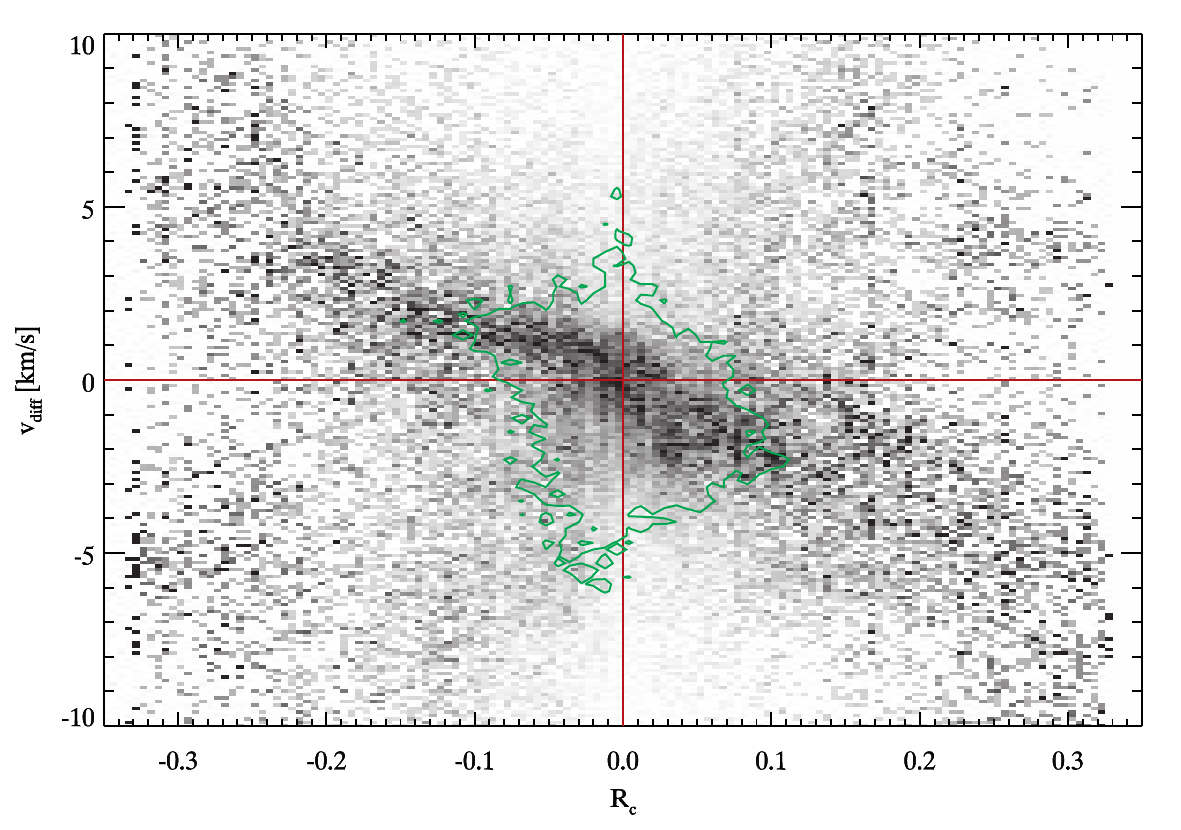} 
  \caption[]{\label{asym}
PDF of the velocity difference between the core and the peak formation heights (Equation~\ref{vd}) as
a function of the line asymmetry (Equation~\ref{ir}). Red lines indicate $R_c\is 0$ and $v_{\rm diff}\is 0$. 
}
\end{figure}

Figure~\ref{asym} shows how this asymmetry measure is correlated with the velocity difference defined above.
Blue peak stronger than the red peak (positive $R_c$) is correlated with a downflow of matter (relative to
the velocity at the peak formation height) above the peak
formation height (negative $v_{\rm diff}$).

\subsection{Line shifts}

Velocities along the line of sight in the atmosphere will cause a Doppler shift of the atomic absorption profile.
If the velocity does not vary too much across the formation region of the intensity, we also get a corresponding
Doppler shift of the intensity profile. We will here look at the correlations between the velocities in the
atmosphere and the line core Doppler shift, the possible usage of the difference in Doppler shift between 
the two main \cii\ lines to diagnose velocity gradients and finally the usage of a Gaussian fit to the whole
line profile as a velocity diagnostic.

\subsubsection{Doppler shift of the line core}

\begin{figure}[hbtp]
  \centering
  \includegraphics[width=\columnwidth]{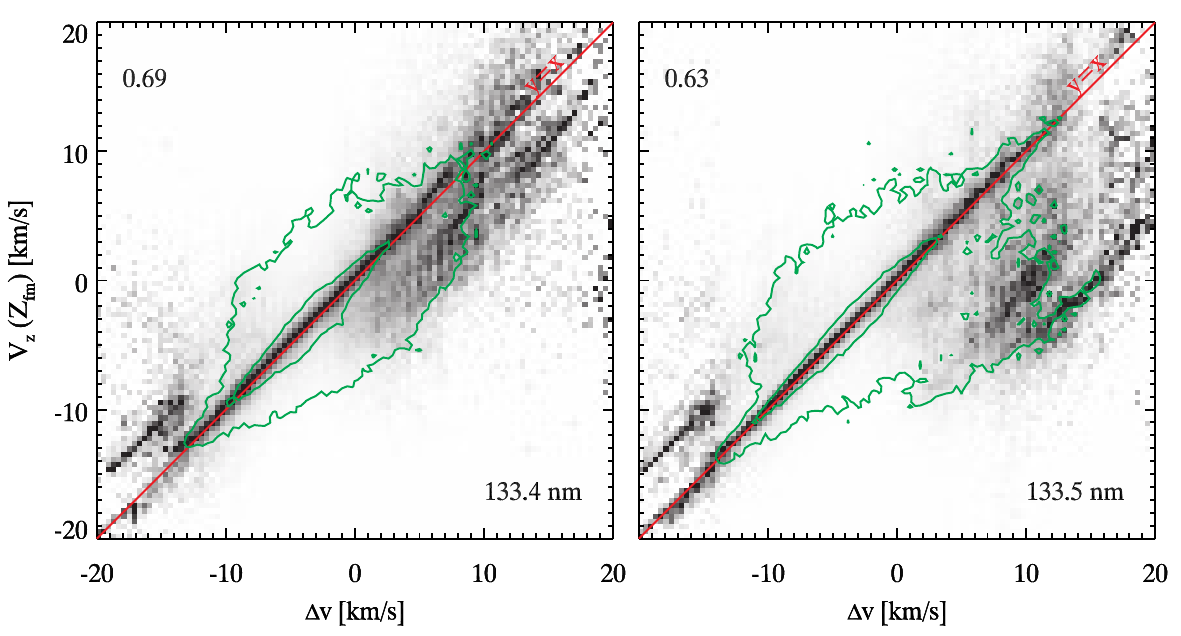} 
  \caption[]{\label{corr_vz}
PDF of the line-of-sight velocity as a function of Doppler shift of the line core. 
The velocity is at the intensity formation height (Equation~\ref{zfm}) of the core wavelength.
Positive velocity is upflow (blue shift).
The line core is defined as in Section~\ref{sec:linecore}.
The green contours encompass 50\% and 90\% of all points. 
Each column in the panels is scaled to maximum contrast to increase visibility. 
The Pearson correlation coefficient is given in the upper left corner. 
The red line denotes the line y=x. 
}
\end{figure}

By choosing the line core as defined in Section~\ref{sec:linecore} we
have the hope of getting a diagnostic of the velocity in the atmosphere close to the transition region. 
Figure~\ref{corr_vz} shows the correlation of the Doppler shift of the line core and the line-of-sight 
velocity at the formation height of the intensity at that wavelength ($Z_{\rm fm}$ in Equation~\ref{zfm}).

Figure~\ref{corr_vz} shows there is a good correlation between the vertical component of the velocity at the 
formation height and the Doppler shift of the line core. 
What appears as a bad correlation at the extreme ranges only concerns a small number of
cases. The green contour and the Pearson correlation coefficient show that more than 50\% of the points
have a tight correlation. 

An interesting feature can be seen in the right panel of Figure~\ref{corr_vz}: a dark patch on the
blue-ward side (positive $\Delta v$) for the \ciib\ line along a line parallel to the y=x line. 
These pixels represent columns where the core finding algorithm has found the blend at 133.566~nm
instead of the main component. The blend is at a wavelength corresponding to a blueshift of 9 km s$^{-1}$.
The other bands on either side of the y=x line (and on the blue side of the band caused by the blend
at 133.566~nm) for large absolute values of $\Delta v$ represents
columns where the core-finding algorithm has picked out the blue or red peak as the line core. This may
happen when there are large velocity gradients smearing out the central depression such that what is
really a double-peak profile appears as a very asymmetric single-peak profile.

\begin{figure}[hbtp]
  \centering
  \includegraphics[width=\columnwidth]{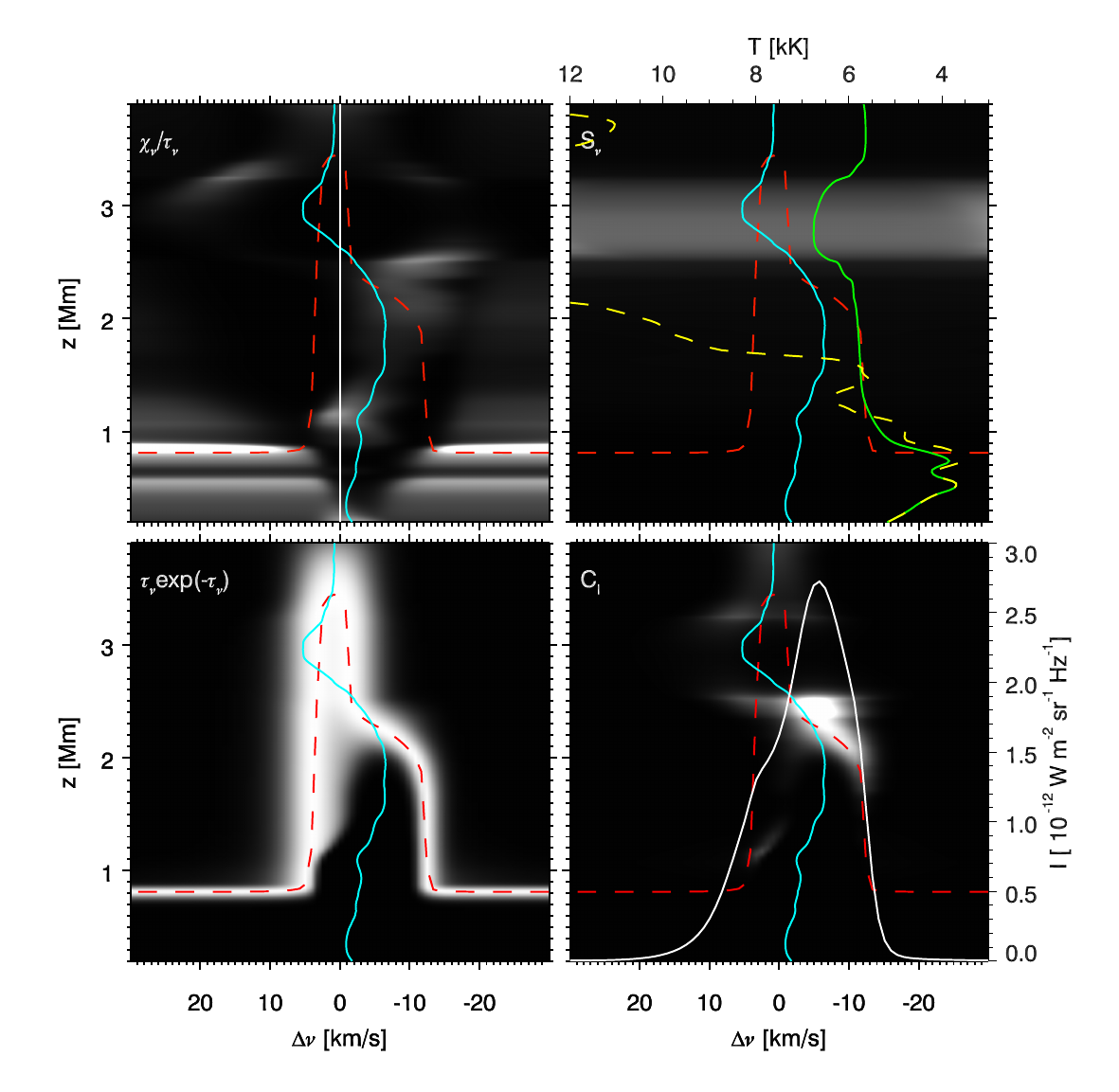} 
  \caption[]{\label{iform_250_182_1334}
Contribution function to the intensity of the \ciia\ line showing an example of an outlier. 
   Each grey-scale image shows the quantity specified in
    its top-left corner as function of frequency from line center (in
    Doppler shift units) and atmospheric height $z$.  Multiplication of
    the first three produces the contribution function to specific intensity shown in
    the lower right panel. A $\tau_\nu\is1$ curve ({red dashed}) and the
    vertical velocity ({blue solid}, positive is upflow) are
    shown in each panel, with a $v_z \is 0$ line in the upper left
    panel for reference.  The image in the upper-right panel is the total source function with lines showing the
    Planck function ({yellow dashed}) and the line source function ({green solid}) in radiation temperature 
    units specified along the
    top. The lower-right panel also contains the emergent intensity
    profile ({white solid}), with the scale along the
    right-hand side.
}
\end{figure}

An example of an outlier is shown in Figure~\ref{iform_250_182_1334}. At this column of the atmosphere,
there is a strong velocity gradient. An upward velocity at 3 Mm height moves the atomic absorption profile to the
blue and a downflow at 2 Mm moves the profile to the red. This creates a $\tau_\nu =1$ (lower left panel) 
graph with a narrow peak at 3.5~Mm and a shoulder towards the red at 2--2.5~Mm. 
The contribution function to intensity is maximum at \edt{2.5~Mm} (lower right panel)
but with a maximum on the red side of the velocity curve because of the gradient term $\chi_\nu/\tau_\nu$
(upper left panel). Our algorithm picks out the maximum intensity wavelength (which has a redshift
of 5 km s$^{-1}$) as the line core (since
this is a single-peak profile) although the wavelength where the optical depth unity is the highest 
is at zero shift. The velocity we measure (5 km s$^{-1}$) is also 2 km s$^{-1}$ larger than the actual
down flow at the formation height at this wavelength.
The strong velocity gradient is the cause of both these effects. With zero velocity we would get a much
narrower intensity profile with two intensity peaks where optical depth unity is coinciding with the 
source function maximum at 2.8~Mm and a central reversal with the intensity formed at the maximum
\edt{height of unit optical depth} of 3.5~Mm. 

\subsubsection{Combining both lines}
\begin{figure}[hbtp]
 \centering
  \includegraphics[width=80mm]{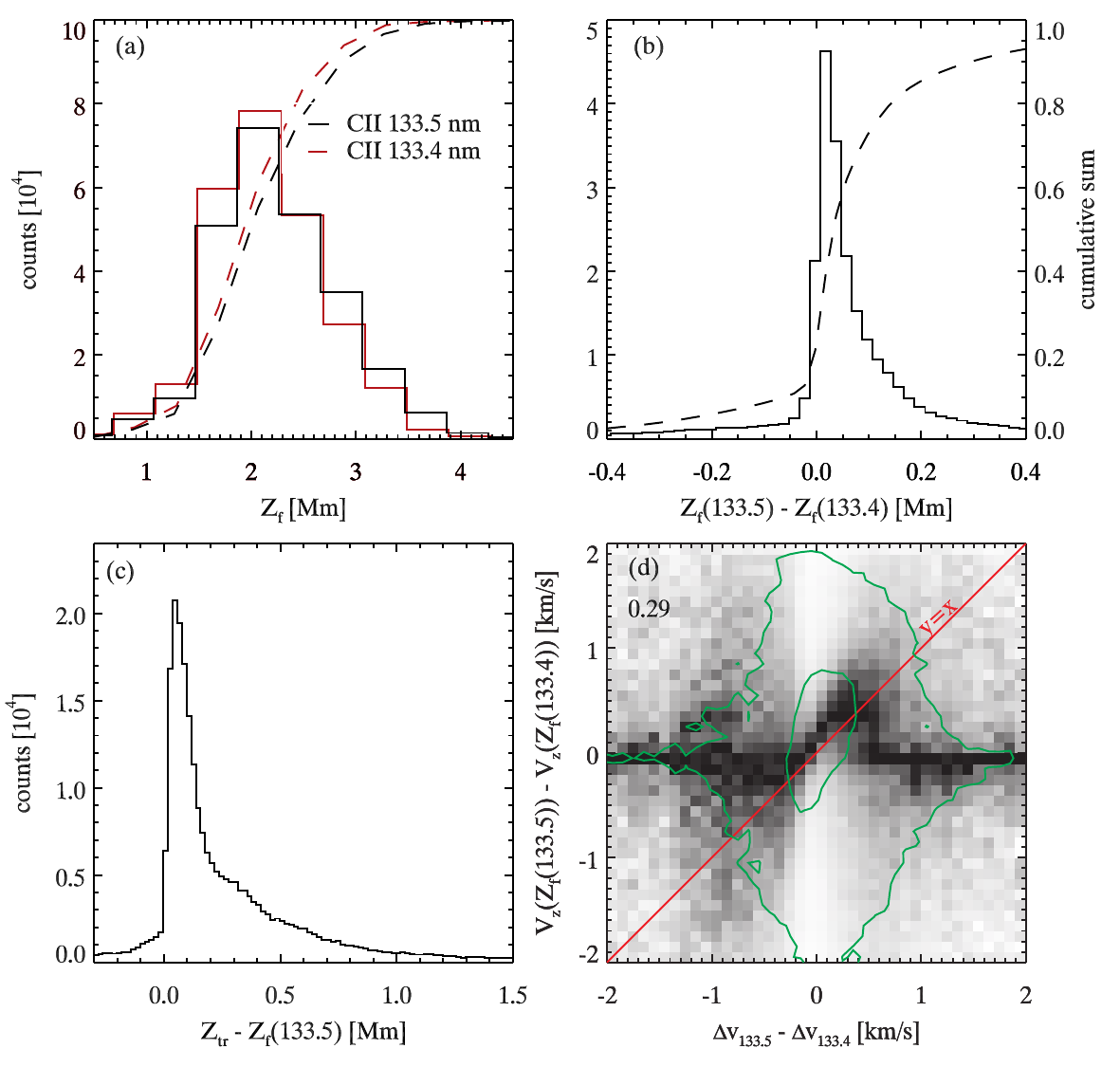}
  \caption[]{\label{hist_zfm}
Different properties of the formation height ($Z_{\rm fm}$ in Equation~\ref{zfm}) of the \ciia\ and \ciib\ line
cores (as found by the core finding algorithm). 
Histogram of formation height for both lines, (panel a, \ciia\ (red) and \ciib\ (black)). 
Histogram of the difference between the formation heights of the lines (panel b). 
The cumulative sum of the histogram is shown with a dashed curve with the scale on the right hand side. 
Histogram of the difference between the transition region height and the formation height of the strongest line
(panel c).
PDF of the difference of the line-of-sight velocity at the formation height 
as a function of the difference in Doppler shift of the line cores (panel d).
Each column is scaled to maximum contrast to increase visibility. 
The Pearson correlation coefficient for the data in the \edt{[-0.5,0.5]} interval in $x$ is given in the upper left corner.
The red line denotes the line $y=x$. 
The green contours encompass 50\% and 90\% of all points.
}
\end{figure}

Here we will discuss the possibility of using the lines together as a diagnostic of velocity gradients. The
\ciib\ line has 1.8 times the opacity of the \ciia\ line and is thus formed higher up in the atmosphere.
Figure~\ref{hist_zfm} shows different properties for the formation of the two lines: histogram of the
formation heights and the difference in formation height, formation height relative to the transition region
and finally the correlation between the difference in Doppler shift and the difference in velocity between
the two formation heights.

The two line cores are formed close to the transition region (Figure~\ref{corr_ztr}) so the histogram in panel (a) of
Figure~\ref{hist_zfm} basically shows the \edt{distribution} of the height of the transition region \edt{over the simulation columns}. The \ciib\ line is formed
slightly higher than the \ciia\ line, the peak of the distribution in panel (b) is at around 30 km. 
The opacity is a factor of 1.8 higher in the \ciib\ line so optical depth unity is always higher by about 0.6 scale heights.
However, the formation height shown in Figure~\ref{hist_zfm} is \edt{weighted by the contribution function,  and differences in the source function between the two lines may give a larger or smaller height difference than the expected 0.6 scale heights}. 
Also\edt{, more importantly,} the core finding
algorithm may identify different wavelengths with respect to the rest wavelength as the core wavelength, especially in
the presence of velocity gradients in the atmosphere and the presence of the blend on the blue side of the \ciib\ line.
These effects account for the small number of points with a negative height difference in panels (b) and (c) and also for 
the tail towards large height differences.
The difference in Doppler shift of the cores of the two lines is correlated with the difference between the
line-of-sight velocities at the \edt{respective} formation heights when the Doppler shift difference is within 0.5 km~$s^{-1}$. 
Larger
differences correspond to problems with the core algorithm and the correlation disappears.

\subsubsection{Single Gaussian fit shift}

Using the line core \edt{Doppler shift} as a diagnostic of atmospheric velocities has the advantage that the line core is formed
over a rather narrow height range compared with the intensity of the whole line. The diagnostic also provides a measure of
the velocity in the uppermost chromosphere. The disadvantage is that sometimes the observationally determined line
core is not really the part of the line \edt{that is formed in the highest region of the atmosphere}. In noisy data there are also challenges in finding the number 
of peaks and the proper line core.

\begin{figure}[hbtp]
  \centering
\hspace{-5mm}
  \includegraphics[width=80mm]{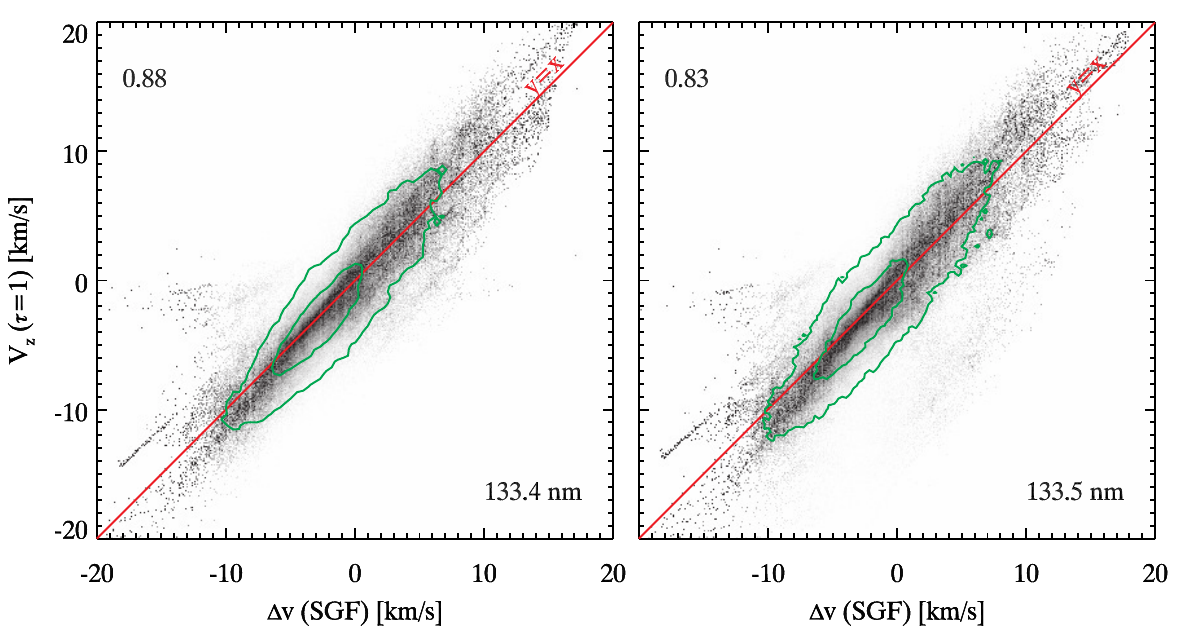} 
  \caption[]{\label{corr_fm_zt1}
PDF of the line-of-sight velocity at the $\tau = 1$ height  as a function of the single Gaussian fit shift
for the \ciia\ line (left panel) and the \ciib\ line (right panel). 
The optical depth unity is taken at the wavelength of the Gaussian shift.
The green contours encompass $50\%$ and $90\% $ of all points. 
Each column in the panels is scaled to maximum contrast to increase visibility. 
The Person correlation coefficient is given in the upper left corner. 
The red line denote the line y=x.
}
\end{figure}

An often used alternative is to use a Gaussian fit to the full line profile. The advantage is that all the points in the profile 
are taken into account, thus minimising the effect of noise. The disadvantage is that we get an influence on the shift from
a much larger height range in the atmosphere. The profile may also deviate substantially from a Gaussian shape for 
a line formed under optically thick conditions or a line with a blend (like the \ciib\ line). For a double-peak profile, one 
could devise a fitting with an emission Gaussian profile fitting the wings and a separate, absorption Gaussian, to fit the core.
The additional fitting parameters demand a good signal-to-noise ratio to be robust. We therefore here test the simpler
procedure of using a single Gaussian fit even for clearly non-Gaussian profiles (like double-peak ones) as a measure of
the shift of the line. To overcome the bias towards the blue from the blend in the \ciib\ line we use the intensity
averaged wavelength between the two components as the reference wavelength (133.57006~nm in this simulation) rather than the
wavelength of the stronger component. 
Figure~\ref{corr_fm_zt1} shows that the shift from a single Gaussian fit is a good diagnostic of the atmospheric velocity
at the \edt{height of unit optical depth}.

An alternative to the single Gaussian fit is to use the first moment of the intensity with respect to wavelength. 
Again it is important to use the intensity weighted 
wavelength for the reference wavelength of the \ciib\ line. Furthermore, the first moment measure is more sensitive
to noisy data in the wings of the profile than the single Gaussian fit.

\subsection{Line width}

The width of the \cii\ lines depends both on the width of the opacity profile  (which is affected by the thermal 
width set by the temperature at the height of formation and by non-thermal motions) and 
by the variation of the source function between the height of formation of the continuum and the core of the line (see Paper~I 
for a detailed discussion). We can characterise the width by several measures; the most common being 
the full-width-at-half-maximum of the intensity profile, $W_{\rm FWHM}$, 
the standard deviation of a single Gaussian fit, $\sigma$, 
the half width at $1/e$ of the maximum intensity, $\Delta V_D$ (which is related to the most probable speed) 
and the second moment of the intensity with respect to wavelength relative to the first moment shift, $W_2$.

For a Gaussian intensity profile we have the following relations between the four width measures:
\begin{eqnarray}
\Delta V_D&=&\sqrt{2}\sigma \label{eq:dvd}\\
W_{\rm FWHM}&=&2\sqrt{\ln 2}\Delta V_D \label{eq:fwhm}\\
W_2&=&\sigma \label{eq:w2}.
\end{eqnarray}

\begin{figure}[hbtp]
  \centering
  \includegraphics[width=\columnwidth]{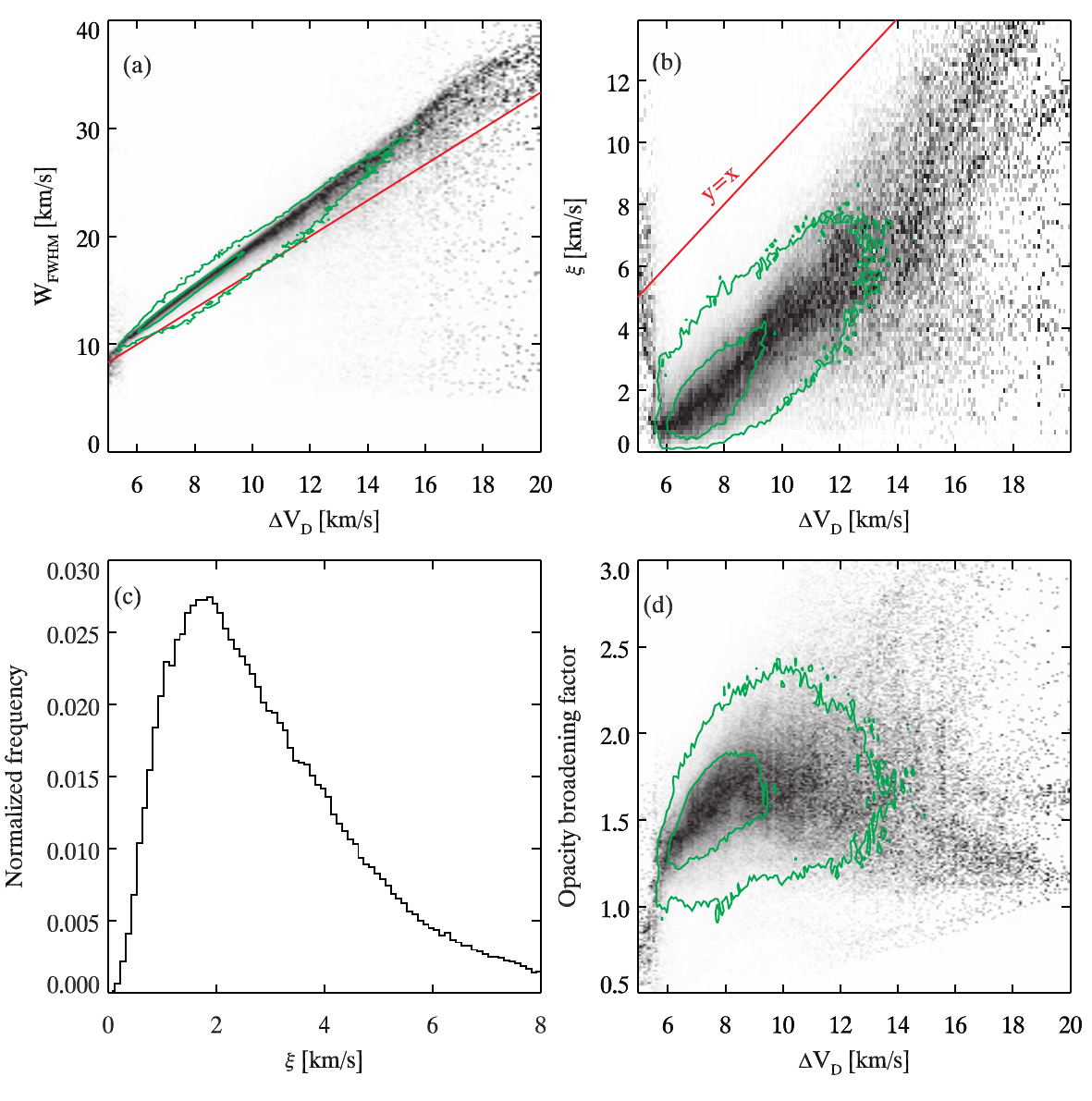} 
  \caption[]{\label{width_sm}
Relations between line-width characteristics:
Panel (a): PDF  of $W_{\rm FWHM}$ as a function of $\Delta V_D$  of a single Gaussian fit. The red line shows the relation
for a Gaussian intensity profile. 
Panel (b): PDF of the non-thermal velocity, $\xi$, as a function of $\Delta V_D$ of a single Gaussian fit.
Panel (c): Histogram of non-thermal velocities.
Panel (d): PDF of opacity broadening factor as function of $\Delta V_D$ of a single Gaussian fit.
The non-thermal velocity is taken as $\sqrt{2}$ times the root-mean-square of the line-of-sight velocities in the
height range between optical depth unity in the continuum and the line core.
The green contours encompass 50\% and 90\% of all points. 
Each column in the PDF panels is scaled to maximum contrast to increase visibility. 
All correlations are for the \ciia\ line.
}.
\end{figure}

Figure~\ref{width_sm} shows relations between various line-width characteristics for the \ciia\ line. 
The upper left panel shows that the full-width-at-half-maximum ($W_{\rm  FWHM}$) is larger than equation~\ref{eq:fwhm}
applied to the 1/e width ($\Delta V_D$) of a single Gaussian fit to the line profile 
(most points are above the red line). This implies that the line profiles in general have a "flatter top" than a Gaussian
profile. This is of course true for the double-peak profiles that tend to be the broader ones. There are some points
in the lower right part of the panel with small $W_{\rm  FWHM}$ and large $\Delta V_D$. These are very asymmetric
double-peak profiles where one peak is less than half the intensity of the other such that the $W_{\rm  FWHM}$ 
corresponds to the width of only the stronger peak while the single Gaussian fit is still affected by the full profile.
The upper right panel of Figure~\ref{width_sm} shows the relation between the measured width of the profile (given
as $\sqrt{2}\sigma$ of a single Gaussian fit) and the non-thermal velocity in the line-forming region. The non-thermal 
velocity is defined as $\sqrt{2}$ times the root-mean-square of the line-of-sight velocities in the
height range between optical depth unity in the continuum and the line core. As in the other correlation figures we have
the observable quantity on the x-axis although the functional relationship is rather $x=f(y)$. The observed width is
well correlated with the non-thermal velocities in the atmosphere\edt{, except for $\Delta V_D<6$~km~s$^{-1}$. These
outliers correspond to profiles with a dominant optically thin component. 
The width is then dominated by the thermal width and the non-thermal velocities in a rather narrow formation region
rather than the in the full height range between the heights of optical depth unity in the continuum and the line core.}
The lower left panel of Figure~\ref{width_sm} shows a histogram of the non-thermal velocities. 
The maximum of the distribution is close to 2~km~s$^{-1}$, which is a rather small value. This will
be further discussed in Section~\ref{sec:discussion}.
The lower right panel of Figure~\ref{width_sm} shows the "opacity broadening factor" defined from

\begin{equation}
{\rm Opacity~broadening~factor}={W_{\rm FWHM}\over2\sqrt{\ln 2}\sqrt{{2kT\over m} + \xi^2}}
\end{equation}

where $T$ is the weighted average of the temperature over the line-forming region using the contribution function to
total intensity as weighting function, $m$ is the mass of carbon and $\xi$ is the non-thermal velocity defined above.

The non-thermal velocities as defined here  give a contribution both to the width
of the opacity profile (when the length scale of the velocity variations is small compared with the line-forming length scale;
classical micro turbulence) and to shifts of the profile (for large length scales; classical \edt{macroturbulence}). We can thus
expect the opacity broadening as defined here to be lower than the effect discussed in Paper~I and it may even be
below one (which is the case for only very few points in Figure~\ref{width_sm}). The opacity broadening is mainly around
a factor of 1.5, is smallest for the narrowest profiles (otherwise they wouldn't be that narrow) and bifurcates into large and 
small values for the broadest profiles. 

\section{Comparison with other  {\it IRIS} spectral lines.} \label{sec:compare_iris}

\begin{figure*}[hbtp]
  \centering
  \includegraphics[width=\textwidth]{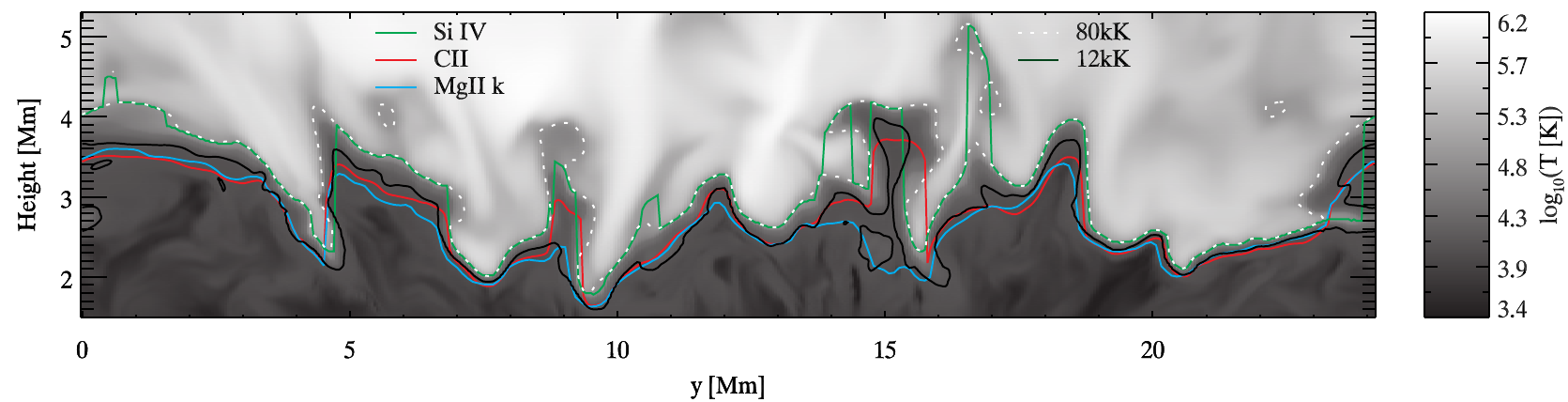} 
  \caption[]{\label{cmgsi}
Formation height of the  \ciib\ line, the \mgiik\ line and the \siiv\  139.3$\,$nm line 
for the snapshot cutout at $x\is 12\,$Mm shown in Figure~\ref{t_nh_cm}. 
Temperature on a logarithmic scale is shown as a greyscale image with contours at 80~kK (white dotted) and
12~kK (black). Maximum $\tau\is 1$ heights are shown for the \ciib\ line (red) and for the \mgiik\  line (blue) and
the maximum emissivity height is shown for the \siiv\  139.3$\,$nm line (green). 
}
\end{figure*}

In Figure~\ref{cmgsi} we compare the formation heights of the  \ciib\ line, the \mgiik\ line and the 
\siiv\  139.3~nm line for the snapshot cutout at $x\is 12\,$Mm shown in Figure~\ref{t_nh_cm}. 
As formation height we use the maximum $\tau\is 1$ height for the \cii\ and \mgiik\ lines and for \siiv\ we use the height 
of maximum emissivity as calculated from CHIANTI. 
This comparison shows that the \siiv\ line normally has the highest formation height at a temperature around 80~kK
with the \cii\ line and \mgiik\ line formed lower. The \cii\ line is at some locations formed at a lower height than
the \mgiik\ line (\eg\ for $y\is 0-4$~Mm) but often higher. 

\begin{figure}[hbtp]
  \centering
	  \includegraphics[width=\columnwidth]{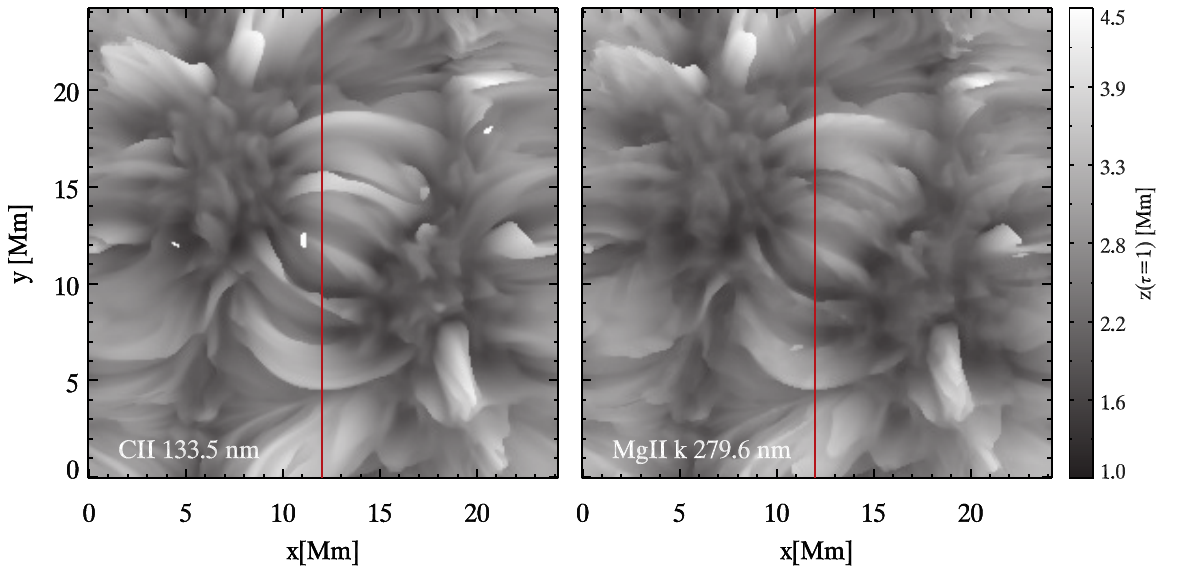} 
  \caption[]{\label{fig:zcmg}
Maximum \edt{height of unit optical depth} over the line profile for  the \ciib\  (left) and \mgiik\  (right) lines.
The location of the cutout shown in Figure~\ref{cmgsi} is shown with a red line at $x\is 12\,$Mm.
}
\end{figure}

We show the formation heights of the  \ciib\ and the \mgiik\  lines in the whole box in Figure~\ref{fig:zcmg}.
The small patches of very high formation height for the \ciia\ line (\eg\ at $(x,y)\is(11,12)\,$Mm)
are caused by cooler pockets of plasma at large heights that have enough density to place $\tau\is 1$ there.
However, the temperature is not low enough in these bubbles for \mgiik\ to reach optical depth unity. From the
figure it is clear that the \ciib\ and \mgiik\ formation heights show very similar patterns but that the
\cii\ line is formed higher up in the fibrils in the central part of the simulation domain (\eg\ at $y\is 15\,$Mm), 
see also Figure~\ref{cmgsi}.

The line opacity per unit mass is given by
\begin{equation}\label{eq:kappanu}
\kappa^l_\nu={\pi e^2\over m_e c^2} n_l f_{lu}\phi_\nu {1\over\rho}
\end{equation}

where $e$ is the electron charge (in e.s.u), $m_e$ is the electron mass, $c$ is the speed of light, $n_l$ is the
population density of the lower level, $f_{lu}$ is the absorption oscillator strength, $\phi_\nu$ is the atomic
absorption profile and $\rho$ is the mass density. We have here neglected stimulated emission. Assuming a
Doppler absorption profile (which is a good approximation at the line core forming region for these lines) we get
for the opacity at line center for a spectral line of a singly ionized state:

\begin{equation}\label{eq:kappanu0}
\kappa^l_{\nu_0}=0.02654 {n_l\over N_{\rm II}}{N_{\rm II}\over N_{\rm el}}{N_{\rm el}\over N_{\rm H}}
{N_{\rm H}\over \rho} f_{lu} {1\over \sqrt{\pi}\Delta\nu_D}
\end{equation}

where $N_{\rm II}$  is the number density of the singly ionized state, $N_{\rm el}$ is the number density of the 
element, $N_{\rm H}$ is the number density of hydrogen particles and $\Delta\nu_D$ is the $1/e$ width of the
atomic absorption Doppler profile (in frequency units). The four population ratios are the fraction of the singly ionized
state that is in the lower level of the transition (4/6 for the \ciib\ line, 1 for the \mgiik\ line), the ionization
fraction of the singly ionized state, the abundance of the element and the number of hydrogen particles per
unit mass (a constant only dependent on the abundances), respectively. The $1/e$ width is given by
\begin{equation}
\Delta\nu_D={1\over\lambda_0}\sqrt{{2 k T\over m} + \xi^2}
\end{equation}

The abundance of carbon is a factor of 6.8 larger than that of magnesium
\citep{2009ARA&A..47..481A} while the oscillator strength is a factor of 5.2 higher for \mgiik. The thermal broadening
is a factor of 1.4 larger for carbon (due to a factor of two lower mass). Inserting these numbers into
Equation~\ref{eq:kappanu0}, and \edt{assuming identical ionization fractions} for a moment, we find an opacity
a factor of  3.4 larger for the \mgiik\ line if thermal broadening dominates
$\Delta\nu_D$ and a factor of 2.4 larger if the non-thermal broadening dominates. Integrating this equation
from the corona and downwards, the \edt{ionization fraction becomes critical}. When the temperature drops below 50~kK
we start to get substantial amounts of \cii\ (10\% ionization fraction at this temperature, see Paper~I) while we still
have negligible amounts of \mgii. The optical depth starts to increase for the \ciib\ line while the \mgiik\ line
\edt{still has} negligible optical depth. At a temperature of 16~kK we have 10\% \mgii\ \citep{2012A&A...539A..39C}
and almost 100\% \cii. At a temperature lower than 14~kK, the optical depth builds up faster in the \mgiik\ line than in the
\ciib\ line because of the \edt{larger} opacity factor estimated above and because carbon gets neutral below 8~kK while magnesium stays
mostly in the singly ionized state to much lower temperatures . The \ciib\ line reaches optical depth unity
at a greater height than the \mgiik\ line if there is a sufficient amount of
material in the 14--50~kK temperature range, otherwise the \mgiik\ line is formed higher. Both cases are present
in our \bifrost\ simulation as is evident from Figures~\ref{cmgsi}--\ref{fig:zcmg}. Velocity gradients
will complicate this picture since the \mgiik\ line has a smaller thermal width and the opacity is therefore more
sensitive to velocity gradients. This is seen in Figure~\ref{fig:zcmg} as locally smoother $\tau\is 1$ surfaces for the 
\ciib\ line than for the \mgiik\ line.

\section{Comparison with Observations}\label{sec:observations}

The main purpose of this paper is to explore the diagnostic potential of the \ciia\ and \ciib\ lines. For that purpose
we have used a \bifrost\ simulation cube to map how atmospheric parameters are encoded in observable quantities.
A comparison between the synthetic observables and observations will furthermore give information on what might
be missing in the simulations for a realistic description of the solar atmosphere. It is important to stress, however, that
we are not dependent on an
accurate match between the synthetic observables and observations for the derived mapping of atmospheric parameters
and observables to be relevant; it is enough that the simulation cube spans a relevant parameter range.

For this comparison we use observations from the \textit{IRIS} satellite taken on 2014 February 25 at 20:50 UT. 
This is a very large dense raster with 400 raster steps of $0.35''$ with a spatial sampling of $0.16"$ along the slit. 
The exposure time was 30s for each raster step with 31.7s step cadence and there is no binning in space or wavelength.
The raster covers a field-of-view of $141'' \times 174''$  centered at $(x,y)\is(-73'',75'')$ and it took 3.5 hours to complete.
The area observed corresponds to quiet Sun.

\begin{figure}[hbtp]
  \centering
  \includegraphics[width=\columnwidth]{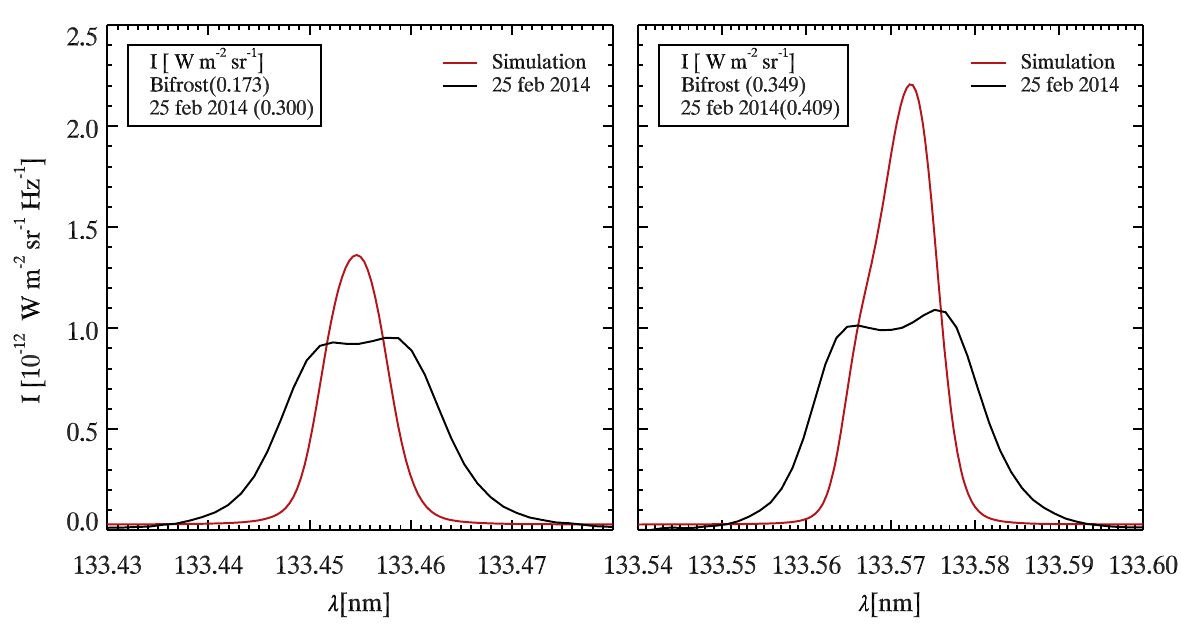} 
  \caption[]{\label{intm}
\cii\ average line profiles from a quiet sun internetwork region observed on 25 february 2014 at 20:50 UT 
with the \textit{IRIS} satellite (black) compared with the average spectral profiles from our \textit{Bifrost} simulation
snapshot (red).  The simulation profiles have been convolved with the {\it IRIS} spectral resolution.
The total intensity is given in the upper left corner. 
}
\end{figure}

We have used \textit{IRIS} calibrated level 2 and level 3 data, details of the data reduction are given in \citet{2014SoPh..289.2733D}. 
The mean intensity profile of the quiet sun internetwork region is shown in Figure~\ref{intm}. 
The total intensity summed over both lines is 0.71 W~m$^{-2}$~sr$^{-1} $ in the observations
which should be compared with 0.52 W~m$^{-2}$~sr$^{-1} $ that we get from the simulation cube and
0.42 and 0.87 W~m$^{-2}$~sr$^{-1}$ from the two SUMER datasets reported by \citet{2003ApJ...597.1158J}.

While the total intensity of the two lines is similar between the simulation and the observations, there are some major
differences. The simulations give a single-peak profile while the observations show a 
double-peak, much broader profile. The ratio between the peak intensities is on average 1.4--1.6 in the simulation and 1.1--1.2 in the
observations. This discrepancy is there not only for the mean profile but also for individual profiles. About 40\% of the observed \ciia\
quiet sun profiles are single-peak  while we have 75\% single-peak profiles in
the simulation (Figure~\ref{npeak}). 
The mean profiles in the observations have an asymmetry with the red peak stronger than the blue peak.
This is opposite of what is the case for the \mgii\ h \& k profiles \citep{2013ApJ...772...90L}.
We speculate that this is caused by  \edt{the fact that the formation of the \cii\ peak happens} mostly above the 
location where the gas pressure is equal to the magnetic pressure ($\beta\is 1$ surface) in the internetwork while the \mgii\ h \& k 
peaks are formed below. The \mgii\ h \& k lines therefore get enhanced blue peaks from acoustic shocks and the
\cii\ lines do not. This is consistent with the visibility of internetwork chromospheric bright grains observed with
{\it IRIS} \citep{2015ApJ...803...44M}.

\begin{figure}[hbtp]
  \centering
  \includegraphics[width=\columnwidth]{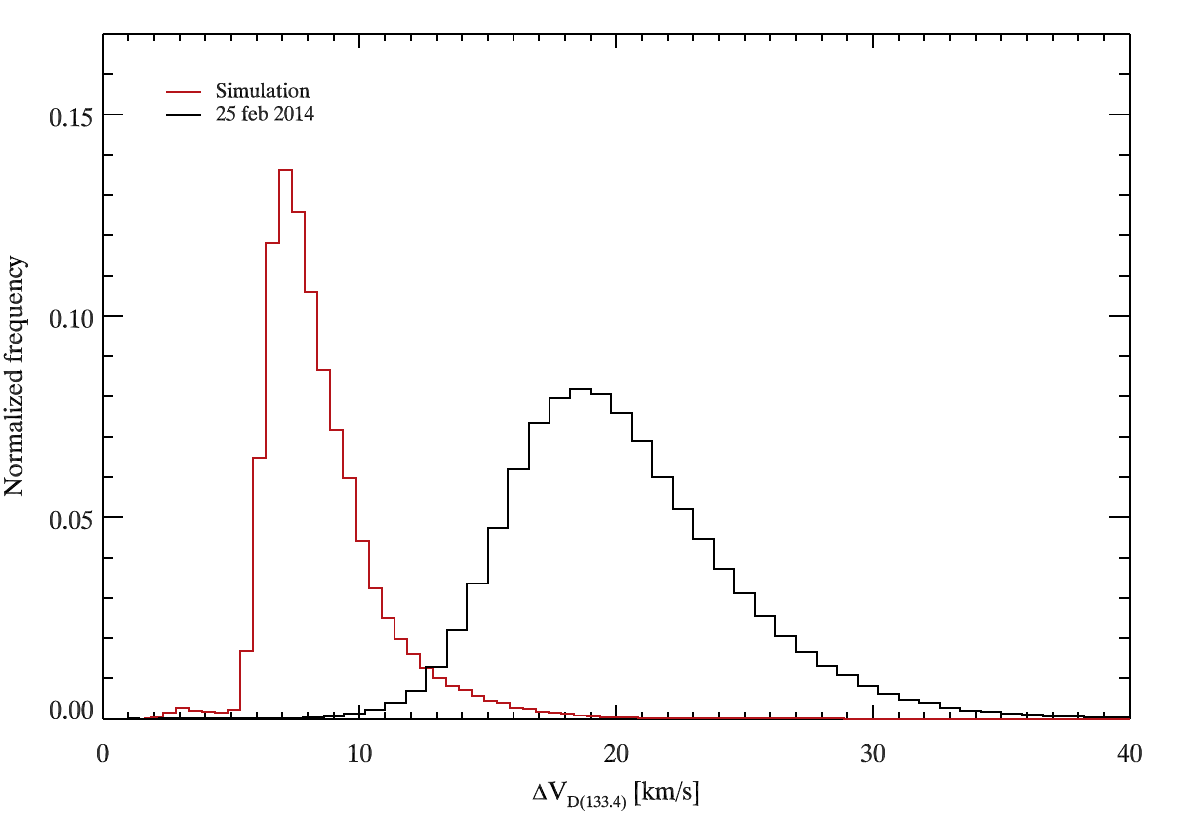} 
  \caption[]{\label{simobs}
Histogram of $\Delta V_D$ for a single Gaussian fit for the \ciia\ line in the simulation (red) and quiet solar 
internetwork region observation (black).  The {\it IRIS} instrumental broadening has been removed from the observed values. 
}
\end{figure}

Figure~\ref{simobs} gives the distribution of the profile widths in the simulation and
in the quiet Sun observational dataset for the \ciia\ line. The average width in the observations is 21~km~s$^{-1}$, 
more than twice the average width of 9~km~s$^{-1}$ we have in the simulation. However, the simulations show a similar
{\em range} in widths with more than 200 profiles wider than 24~km~s$^{-1}$. We discuss possible reasons for the
discrepancies between the simulations and the observations in Section~\ref{sec:discussion}.

\section{Discussion and Conclusions}\label{sec:discussion}

We have studied the diagnostic potential of the \ciia\ and the \ciib\ lines that are among the strongest
lines in the FUV passbands of the {\it IRIS} spacecraft.

We found that local maxima in the source function between the formation height of the continuum and the line core
can give rise to a variety of profile shapes. In the simulation, single-peak emission profiles dominate but there are also
double and multiple peak profiles. Velocity gradients alter the shape expected from just the source function variation.
We introduced an observational definition of the line core by taking the position of the central peak for profiles with
an odd number of peaks and the position of the central depression for profiles with an even number of peaks. This
definition in most cases gives a good approximation to the theoretical definition of the line core  (the wavelength
where the optical depth unity point is the highest). However, in the presence of strong velocity gradients we may get
a filling in of the central reversal and incorrectly assign one of the peaks as the line core wavelength.
The contribution to the line core intensity has its maximum close to the transition region (defined as being the highest height where
the temperature is below 30~kK). The contribution function to intensity on a logarithmic optical depth scale shows that
the dominant formation is optically thick with occasionally an optically thin component when the source function increases
very strongly into the transition region.

We inspected various relations between atmospheric properties and observables. We found that the low intensity
profiles tend to have a higher formation height. This means that the source function is lower \edt{at the formation height} 
when the transition
region is at a greater height.
The intensity at the line core is only weakly correlated with the temperature at the formation height. Normally, the source
function has decoupled from the Planck function at this height such that the actual temperature at the formation 
height is about twice the radiation temperature of the line core intensity. There is a large scatter and the relation
is of limited practical use.

The intensity ratio between the \ciib\ and the \ciia\ lines is 1.8 in the optically thin case and can be any value (including the
optically thin value of 1.8) in the optically 
thick case depending on the ratio of the source functions of the two lines. This means that a value different from 1.8 
shows optically thick formation while a value of 1.8 does not prove optically thin conditions.
The ratio of the peak intensities in the \ciib\ lines in the \bifrost\ simulation snapshot is on average 1.4--1.7 with the lower value for double-peak profiles. 
Double-peak profiles often show an asymmetry with one peak brighter than the other. This asymmetry is correlated 
with the velocity gradient that exists between the formation height of the peaks and the line core. Blue peak stronger
than the red peak means the atmosphere has a downflow at the core formation height relative to the
motion at the formation height of the peaks.

The line core Doppler shift is well correlated with the line-of-sight velocity at the formation height. Due to the 
difficulties in observationally identifying the line core, there is substantial scatter in the relation. The blend on the
blueward side of the \ciib\ line also cause occasional misidentifications.
The \ciia\ line is formed some 30~km below the \ciib\ line. The difference in Doppler shift between
the two lines is correlated with the velocity gradient in this height interval.
A single Gaussian fit to the whole profile is well correlated with the velocity at unity optical depth. Such a fit is easier
to make in the presence of noise than using our core-finding algorithm that works best for observations with high
signal-to-noise.

The line width is well correlated with non-thermal velocities between the formation heights of the continuum and
the line core. In addition to the width of the atomic absorption profile (which is affected by thermal and non-thermal
motions) we get an additional broadening due to the optically thick line formation. This additional broadening, often
called "opacity broadening", is an additional factor in the range 1.2--2 but sometimes reaching 4. The factor depends
on the behaviour of the source function between the formation heights of the continuum and the line core. For
single-peak profiles with a source function that rises steeply into the transition region, the opacity broadening is small 
and for source functions that rise rapidly in the lower chromosphere and then flatten out or decrease, the opacity
broadening is large.

We compared the formation height of the \ciib\ line with that of \mgiik\ and \siiv~139.3~nm. The \siiv\ line
is formed the highest around 80~kK temperature while the \cii\ and \mgiik\ lines are formed 
lower at similar heights below the transition
region. The relative formation height of the two lines depend on the amount of matter in the 14--50~kK temperature
range. With significant amounts of matter in this temperature range where magnesium is still more than singly ionized,
the \ciib\ line is formed above the \mgiik\ line but with less material in this temperature range we have 
the opposite situation.

We compared the simulation results with recent observations from {\it IRIS}. The total intensity of the mean profiles from 
the simulations is in general agreement with the observed values but the observed profiles are a factor of 2.5
wider than the ones in the simulation. \edt{Also, the observed intensity ratio between the \ciib\ and \ciia\ lines
ranges between 1.1 and 1.2, compared to the range 1.4--1.7 found in our
simulations}. The mean profile from the simulations is single peaked while in observations
of the quiet Sun, the mean profile is double peaked. There is also a larger proportion of single-peak profiles in the simulations
than in the observations.

Double-peak profiles come from the existence of a local source function maximum between the formation heights of the
continuum and the line core caused by a temperature rise and sufficient coupling between the Planck function and
the source function. This means that we have too low temperatures in the low-mid chromosphere in the simulation compared
with what observations of the quiet Sun reveal. Increasing the number of double-peak profiles through increased
heating in the lower chromosphere in the simulation would also increase
the width through increased opacity broadening. In addition, this comparison shows that we have too small non-thermal 
velocities in the middle chromosphere in the simulations. 

All the correlations have been derived with the full spectral and spatial resolution of the \bifrost\ simulation. At {\it IRIS} resolution
we still expect most of the results to be valid. Convolving our intensity profiles with the {\it IRIS} resolution decreases the
proportion of double-peak \ciib\ profiles from 23\% to 9\%\edt{, which further increases the disproportion of single-peak
over double-peak profiles in our simulations compared to observations. However, we} expect the effect on the solar profiles to be much smaller since 
our synthetic profiles are too narrow by a factor of two. The correlation between Doppler shift differences and velocity
differences between the two \cii\ lines was only significant for Doppler shift differences smaller than 0.5~km~s$^{-1}$. Measuring Doppler shifts to that precision with the {\it IRIS} spectral pixels of 2.9~km~s$^{-1}$ is possible but requires high signal-to-noise.

Although there are large differences between the mean properties of the synthetic profiles and the observations, we find
that the simulations cover most of the parameter range shown by the observations of the quiet Sun (although in very
different distributions). We thus believe the derived relations are valid under solar conditions. The mere fact that there
are large differences in the distribution of properties between the synthetic profiles and the observed ones shows that
the \ciia\ and \ciib\ lines are powerful diagnostics of the chromosphere and lower transition region.

\begin{acknowledgements}
The research leading to these results has received funding from the European Research Council under the
 European Union's Seventh Framework Programme (FP7/2007-2013) / ERC grant agreement no 291058.
This research was supported by the Research Council of Norway through the grant ÒSolar Atmospheric 
ModellingÓ and through grants of computing time from the Programme for Supercomputing and
through computing project s1061 from the High End Computing Division of NASA. 
B.D.P. acknowledges support from NASA grants NNX11AN98G and NNM12AB40P and NASA contract NNG09FA40C (IRIS).
\end{acknowledgements}

\bibliographystyle{apj}   

\begin{thebibliography}{}
\expandafter\ifx\csname natexlab\endcsname\relax\def\natexlab#1{#1}\fi

\bibitem[{{Asplund} {et~al.}(2009){Asplund}, {Grevesse}, {Sauval}, \&
  {Scott}}]{2009ARA&A..47..481A}
{Asplund}, M., {Grevesse}, N., {Sauval}, A.~J., \& {Scott}, P. 2009, \araa, 47,
  481

\bibitem[{{Carlsson} {et~al.}(2015){Carlsson}, {Hansteen}, {Gudiksen},
  {Leenaarts}, \& {De Pontieu}}]{chromsim15}
{Carlsson}, M., {Hansteen}, V.~H., {Gudiksen}, B.~V., {Leenaarts}, J., \& {De
  Pontieu}, B. 2015, \aap, (in preparation)

\bibitem[{{Carlsson} \& {Leenaarts}(2012)}]{2012A&A...539A..39C}
{Carlsson}, M., \& {Leenaarts}, J. 2012, \aap, 539, A39

\bibitem[{{de la Cruz Rodr{\'{\i}}guez} {et~al.}(2013){de la Cruz
  Rodr{\'{\i}}guez}, {De Pontieu}, {Carlsson}, \& {Rouppe van der
  Voort}}]{2013ApJ...764L..11D}
{de la Cruz Rodr{\'{\i}}guez}, J., {De Pontieu}, B., {Carlsson}, M., \& {Rouppe
  van der Voort}, L.~H.~M. 2013, \apjl, 764, L11

\bibitem[{{De Pontieu} {et~al.}(2014){De Pontieu}, {Title}, {Lemen}, {Kushner},
  {Akin}, {Allard}, {Berger}, {Boerner}, {Cheung}, {Chou}, {Drake}, {Duncan},
  {Freeland}, {Heyman}, {Hoffman}, {Hurlburt}, {Lindgren}, {Mathur}, {Rehse},
  {Sabolish}, {Seguin}, {Schrijver}, {Tarbell}, {W{\"u}lser}, {Wolfson},
  {Yanari}, {Mudge}, {Nguyen-Phuc}, {Timmons}, {van Bezooijen}, {Weingrod},
  {Brookner}, {Butcher}, {Dougherty}, {Eder}, {Knagenhjelm}, {Larsen},
  {Mansir}, {Phan}, {Boyle}, {Cheimets}, {DeLuca}, {Golub}, {Gates}, {Hertz},
  {McKillop}, {Park}, {Perry}, {Podgorski}, {Reeves}, {Saar}, {Testa}, {Tian},
  {Weber}, {Dunn}, {Eccles}, {Jaeggli}, {Kankelborg}, {Mashburn}, {Pust},
  {Springer}, {Carvalho}, {Kleint}, {Marmie}, {Mazmanian}, {Pereira}, {Sawyer},
  {Strong}, {Worden}, {Carlsson}, {Hansteen}, {Leenaarts}, {Wiesmann},
  {Aloise}, {Chu}, {Bush}, {Scherrer}, {Brekke}, {Martinez-Sykora}, {Lites},
  {McIntosh}, {Uitenbroek}, {Okamoto}, {Gummin}, {Auker}, {Jerram}, {Pool}, \&
  {Waltham}}]{2014SoPh..289.2733D}
{De Pontieu}, B., {Title}, A.~M., {Lemen}, J.~R., {et~al.} 2014, \solphys, 289,
  2733

\bibitem[{{Gudiksen} {et~al.}(2011){Gudiksen}, {Carlsson}, {Hansteen}, {Hayek},
  {Leenaarts}, \& {Mart{\'{\i}}nez-Sykora}}]{2011A&A...531A.154G}
{Gudiksen}, B.~V., {Carlsson}, M., {Hansteen}, V.~H., {et~al.} 2011, \aap, 531,
  A154

\bibitem[{Gustafsson(1973)}]{Gustafsson1973}
Gustafsson, B. 1973, Uppsala Astron. Obs. Ann.,, 5

\bibitem[{{Hummer} \& {Voels}(1988)}]{1988A&A...192..279H}
{Hummer}, D.~G., \& {Voels}, S.~A. 1988, \aap, 192, 279

\bibitem[{{Judge} {et~al.}(2003){Judge}, {Carlsson}, \&
  {Stein}}]{2003ApJ...597.1158J}
{Judge}, P.~G., {Carlsson}, M., \& {Stein}, R.~F. 2003, \apj, 597, 1158

\bibitem[{{Leenaarts} \& {Carlsson}(2009)}]{2009ASPC..415...87L}
{Leenaarts}, J., \& {Carlsson}, M. 2009, in Astronomical Society of the Pacific
  Conference Series, Vol. 415, The Second Hinode Science Meeting: Beyond
  Discovery-Toward Understanding, ed. {B.~Lites, M.~Cheung, T.~Magara,
  J.~Mariska, \& K.~Reeves}, 87

\bibitem[{{Leenaarts} {et~al.}(2012){Leenaarts}, {Carlsson}, \& {Rouppe van der
  Voort}}]{2012ApJ...749..136L}
{Leenaarts}, J., {Carlsson}, M., \& {Rouppe van der Voort}, L. 2012, \apj, 749,
  136

\bibitem[{{Leenaarts} {et~al.}(2013{\natexlab{a}}){Leenaarts}, {Pereira},
  {Carlsson}, {Uitenbroek}, \& {De Pontieu}}]{2013ApJ...772...89L}
{Leenaarts}, J., {Pereira}, T.~M.~D., {Carlsson}, M., {Uitenbroek}, H., \& {De
  Pontieu}, B. 2013{\natexlab{a}}, \apj, 772, 89

\bibitem[{{Leenaarts} {et~al.}(2013{\natexlab{b}}){Leenaarts}, {Pereira},
  {Carlsson}, {Uitenbroek}, \& {De Pontieu}}]{2013ApJ...772...90L}
---. 2013{\natexlab{b}}, \apj, 772, 90

\bibitem[{{Mart{\'{\i}}nez-Sykora} {et~al.}(2015){Mart{\'{\i}}nez-Sykora},
  {Rouppe van der Voort}, {Carlsson}, {De Pontieu}, {Pereira}, {Boerner},
  {Hurlburt}, {Kleint}, {Lemen}, {Tarbell}, {Title}, {Wuelser}, {Hansteen},
  {Golub}, {McKillop}, {Reeves}, {Saar}, {Testa}, {Tian}, {Jaeggli}, \&
  {Kankelborg}}]{2015ApJ...803...44M}
{Mart{\'{\i}}nez-Sykora}, J., {Rouppe van der Voort}, L., {Carlsson}, M.,
  {et~al.} 2015, \apj, 803, 44
\bibitem[{{Ng}(1974)}]{1974JChPh..61.2680N}
{Ng}, K.-C. 1974, \jcp, 61, 2680

\bibitem[{{Olson} {et~al.}(1986){Olson}, {Auer}, \&
  {Buchler}}]{1986JQSRT..35..431O}
{Olson}, G.~L., {Auer}, L.~H., \& {Buchler}, J.~R. 1986, \jqsrt, 35, 431

\bibitem[{{Pereira} {et~al.}(2015){Pereira}, {Carlsson}, {De Pontieu}, \&
  {Hansteen}}]{2015arXiv150401733P}
{Pereira}, T.~M.~D., {Carlsson}, M., {De Pontieu}, B., \& {Hansteen}, V. 2015,
  ArXiv e-prints, arXiv:1504.01733

\bibitem[{{Pereira} {et~al.}(2013){Pereira}, {Leenaarts}, {De Pontieu},
  {Carlsson}, \& {Uitenbroek}}]{2013ApJ...778..143P}
{Pereira}, T.~M.~D., {Leenaarts}, J., {De Pontieu}, B., {Carlsson}, M., \&
  {Uitenbroek}, H. 2013, \apj, 778, 143

\bibitem[{{Rathore} \& {Carlsson}(2015)}]{cii_paper1}
{Rathore}, B., \& {Carlsson}, M. 2015, \apj, (submitted)

\bibitem[{{Rybicki} \& {Hummer}(1991)}]{1991A&A...245..171R}
{Rybicki}, G.~B., \& {Hummer}, D.~G. 1991, \aap, 245, 171

\bibitem[{{Rybicki} \& {Hummer}(1992)}]{1992A&A...262..209R}
---. 1992, \aap, 262, 209

\bibitem[{{{\v S}t{\v e}p{\'a}n} {et~al.}(2012){{\v S}t{\v e}p{\'a}n},
  {Trujillo Bueno}, {Carlsson}, \& {Leenaarts}}]{2012ApJ...758L..43S}
{{\v S}t{\v e}p{\'a}n}, J., {Trujillo Bueno}, J., {Carlsson}, M., \&
  {Leenaarts}, J. 2012, \apjl, 758, L43

\end{thebibliography}

\end{document}